\documentclass[11pt,a4paper]{article}
\ifdefined\XeTeXversion\else\pdfoutput=1\fi
\usepackage{jcappub}
\usepackage[T1]{fontenc}
\usepackage{amsmath,amsfonts,amssymb}
\usepackage{graphicx}
\usepackage{subfig}
\usepackage{xcolor}

\usepackage{array}
\usepackage{booktabs}
\usepackage{tabularx}
\usepackage{adjustbox}
\usepackage{ragged2e}
\usepackage{url}
\usepackage{float}
\usepackage[section]{placeins}
\title{FARSim: a fast RF-chain-aware trigger-screening surrogate for radio detection of ultra-high-energy cosmic rays}

\author[a]{Xin Xu,}
\author[a]{Pengfei Zhang,}
\author[a]{Fufu Yang,}
\author[b,c]{Pengxiong Ma,}
\author[d,e]{Ramesh Koirala,}
\author[d,f]{and Chao Zhang}

\affiliation[a]{School of Electronic Engineering, Xidian University, No.~2 South Taibai Road, Xi'an, China}
\affiliation[b]{Key Laboratory of Dark Matter and Space Astronomy, Purple Mountain Observatory, Chinese Academy of Sciences, No.~10 Yuanhua Road, Nanjing, China}
\affiliation[c]{School of Astronomy and Space Science, University of Science and Technology of China, Hefei 230026, China}
\affiliation[d]{School of Astronomy and Space Science, Nanjing University, 163 Xianlin Road, Nanjing 210023, China}
\affiliation[e]{Space Research Centre, Faculty of Technology, Nepal Academy of Science and Technology (NAST), GPO Box: 3323, Khumaltar, Lalitpur, Nepal}
\affiliation[f]{Key Laboratory of Modern Astronomy and Astrophysics, Nanjing University, 163 Xianlin Road, Nanjing 210023, China}

\emailAdd{xxin\_1@stu.xidian.edu.cn}
\emailAdd{zhangpf@mail.xidian.edu.cn}
\emailAdd{23021110320@stu.xidian.edu.cn}
\emailAdd{mapx@pmo.ac.cn}
\emailAdd{rkoirala.astro@gmail.com}
\emailAdd{chao.zhang@nju.edu.cn}

\abstract{Radio arrays provide a scalable route to detecting extensive air showers from ultra-high-energy cosmic rays, but trigger studies for candidate layouts are expensive when every energy, arrival direction, core position and trigger configuration is evaluated with full radio simulations. We present FARSim, a fast surrogate framework that reuses a reduced library of ZHAireS reference footprints to reconstruct ground-plane radio emission and to estimate trigger-relevant observables. The method combines vector geomagnetic and charge-excess field decomposition, geomagnetic-angle and energy scaling, geometrical projection, contour-based core sampling and event-rate integration. We evaluate an independent validation set of 1999 ZHAireS showers; 941 events satisfy the predefined radial-support and station-count requirements and yield a median event-level relative RMSE of 9.4\% and a median Pearson coefficient $r_{\mathrm{P}}=0.965$ for the reconstructed geomagnetic lateral profiles. We further extend the peak-field surrogate to time-domain electric-field synthesis by combining the predicted three-component peak-field vector with geometry-dependent normalized pulse templates. Propagating these traces through an RF-chain response enables voltage-domain threshold and L1-trigger diagnostics. When evaluated at the true peak amplitude, the time-domain extension reaches a median vector-waveform $R^2$ of 0.986 over 2112 ZHAireS traces held out from pulse-shape model fitting. FARSim is therefore intended as a rapid, physics-informed screening layer for array-layout and trigger studies; absolute exposure predictions and detector commissioning remain the role of full end-to-end simulations.}


\begin{document}
\maketitle

\section{Introduction}

Ultra-high-energy cosmic rays (UHECRs), conventionally defined above $10^{18}\,\mathrm{eV}$, probe the most energetic particle accelerators in the Universe. Their rapidly decreasing flux precludes direct detection at the highest energies, so their properties are inferred from extensive air showers (EAS). The all-particle spectrum contains the knee, second knee, ankle, and a high-energy flux suppression whose interpretation can involve propagation losses, source limits, and composition evolution~\cite{refPDGCosmicRays}. Direct satellite- and balloon-borne measurements remain possible at lower energies, where the flux is higher~\cite{ref1,ref2,ref3}.

When a UHECR interacts with the Earth's atmosphere, it initiates a cascade of secondary particles~\cite{ref16}. Current approaches for EAS detection include optical techniques~\cite{ref4,ref5,refCTA}, particle-detector arrays~\cite{ref6,ref7,ref8,ref9}, and radio measurements~\cite{ref10,ref11,ref12,ref13,ref14,ref15}. The radio emission from EAS arises predominantly from two mechanisms: geomagnetic emission and charge-excess (Askaryan) emission~\cite{ref27,ref28}. In the geomagnetic mechanism, electrons and positrons in the shower are deflected in opposite directions by the Earth's magnetic field, generating coherent radio emission through the transverse current induced by the Lorentz force. In the charge-excess mechanism, a net negative charge excess develops in the shower front, mainly through the entrainment of ionization electrons and positron annihilation, leading to an additional coherent radio-emission component.

Radio-array efficiency depends on both station response and array geometry. For a fixed number of stations, the topology, spacing, and boundary geometry determine the balance among threshold, aperture, and reconstruction quality. Hybrid layouts add a further design degree of freedom by combining regions of different station density.

FARSim is not tied to a particular array geometry or experiment. In principle, its footprint-reconstruction and core-sampling framework can be adapted to ground-based radio arrays for extensive-air-shower detection, including dense, sparse, and hybrid configurations represented by LOFAR, Tunka-Rex, AERA, TREND50, and planned GRAND-related arrays~\cite{ref12,ref13,ref14,ref15,refGRANDlib}. Application to a specific experiment, however, requires replacing or configuring the detector-response module with the corresponding antenna response, analogue and digital RF chain, site-dependent noise and radio-frequency-interference environment, digitization, and local trigger logic. The site-specific geomagnetic field, atmosphere, observation altitude, and terrain must likewise be supplied where they affect the simulated footprint and detector response.

Monte Carlo simulations are typically used to evaluate array performance under different conditions. However, such evaluations require a large number of EAS simulations, with outcomes depending on multiple parameters, including primary energy, zenith and azimuth angles, and the spatial distribution of shower cores within the array footprint~\cite{ref18}. Consequently, assessing the detection efficiency of a fixed layout is computationally expensive and not suitable for large-scale optimization involving many design parameters.

Detailed radio simulations of extensive air showers are commonly performed with full Monte Carlo or microscopic simulation frameworks, such as CoREAS~\cite{refCoREAS} and ZHAireS~\cite{refZHAireS}. These tools provide high-fidelity predictions of the radio emission, but their computational cost becomes a limiting factor when many energies, arrival directions, core positions, and candidate array layouts must be scanned. Several faster, semi-analytical, or modular approaches have therefore been explored to mitigate this burden, including macroscopic modeling of radio emission~\cite{refScholten}, SELFAS-based calculations~\cite{refSELFAS}, NuRadioMC-based detector simulations~\cite{refNuRadioMC}, template-based synthesis methods such as SMIET~\cite{refSMIET}, and RadioMorphing~\cite{ref26}. Motivated by the same practical need---rapid, layout-oriented performance evaluation for large radio arrays~\cite{ref18}---this paper introduces a fast evaluation method for assessing the detection capability of radio arrays.

This work addresses that bottleneck by treating the expensive ZHAireS calculation as a reusable reference footprint rather than as a quantity to be regenerated for every core position, layout, and trigger setting. The resulting method has three elements. First, FARSim reconstructs a ground-plane radio footprint from a reduced reference library by combining vector geomagnetic and charge-excess decomposition, geomagnetic-angle scaling, energy scaling, and geometrical projection. Second, the reconstructed footprint is converted directly into threshold contours and station multiplicities, which makes core-position sampling and event-rate integration inexpensive. Third, the same peak-field surrogate is extended to time-domain electric-field synthesis and RF-chain trigger diagnostics, allowing field-threshold, voltage-threshold, and noisy L1-trigger abstractions to be compared within one framework.

FARSim is therefore distinct from methods with different primary targets. Macroscopic radiation models~\cite{refScholten} formulate the emission from current and charge distributions, SELFAS-based calculations~\cite{refSELFAS} and ZHAireS/CoREAS-like tools target detailed radio-signal prediction, NuRadioMC~\cite{refNuRadioMC} emphasizes detector simulation, template-synthesis approaches such as SMIET~\cite{refSMIET} construct antenna pulses from simulated templates, and RadioMorphing~\cite{ref26} morphs reference radio signals to new shower configurations. FARSim instead asks a narrower trigger-screening question: given a small set of high-fidelity reference footprints, how accurately and how quickly can one estimate trigger regions, multiplicities, and relative event-rate trends for many layouts and trigger abstractions? Its intended role is a rapid screening layer for layout and trigger studies, not a substitute for high-fidelity microscopic radio simulations or full detector validation.

The reference library spans zenith angles from $45^\circ$ to $90^\circ$, but the validated application range ends at $85^\circ$. The principal domain is $60^\circ$--$85^\circ$ for sparse arrays with station spacings of order $500\,\mathrm{m}$. At smaller zenith angles, the compact footprint is poorly sampled by such arrays, and showers developing close to the observation level can exhibit altitude- and geometry-dependent clipping~\cite{refGeoAskaryanModel}. Toward the upper boundary, radio-core recentering mitigates refractive displacement but cannot recover event-dependent amplitude changes or two-dimensional footprint distortions. Source-distance and $X_{\max}$ fluctuations, coherence loss, and additional polarization structure can already affect the emission below $85^\circ$ and generally become more important with increasing inclination. In particular, the coherence and polarization effects depend on the local geomagnetic field and its orientation relative to the shower axis~\cite{refChiche}. The $85^\circ$ limit is therefore a validation boundary for the present configuration, not a universal physical onset; simulations above it are retained only for boundary diagnostics.

The remainder of this paper is organized as follows: section~\ref{sec:trigger_estimation} discusses the difficulties in event-rate estimation and layout optimization; section~\ref{sec:fast_eval} presents the theoretical framework and implementation of the proposed approach, together with validation against dedicated simulations; section~\ref{sec:waveform_rfchain} introduces the time-domain extension and RF-chain trigger diagnostic; and section~\ref{sec:conclusion} summarizes the main conclusions and the scope of applicability of the proposed method.

\section{Triggered-Event-Rate Estimation}
\label{sec:trigger_estimation}

\begin{figure}[!t]
  \centering
  \includegraphics[width=0.95\textwidth]{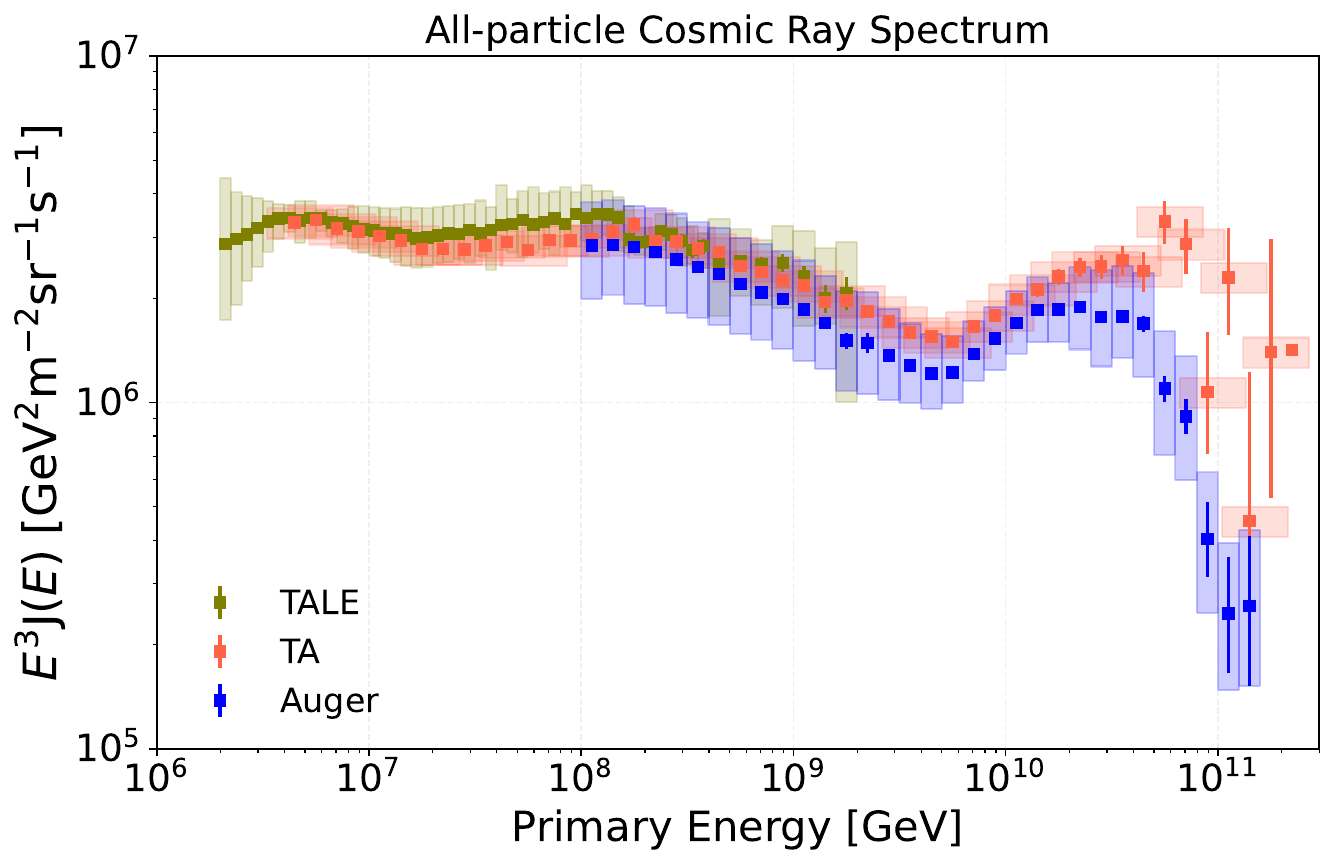}
  \caption{Flux of cosmic rays measured by the TALE~\cite{ref19}, Telescope Array (TA)~\cite{refTA}, and Pierre Auger Observatory~\cite{refAuger}. For orientation, the commonly quoted all-particle landmarks are the knee at a few PeV, the second knee around $100\,\mathrm{PeV}$, the ankle at approximately $4$--$5\,\mathrm{EeV}$, and the flux-suppression or GZK-cutoff region above roughly $40$--$60\,\mathrm{EeV}$~\cite{refPDGCosmicRays}. The TALE spectrum is used here as the baseline flux input for event-rate integration.}
  \label{fig:TALE_flux}
\end{figure}

The differential cosmic-ray flux $J(\mathcal{E})$ is defined per unit energy, solid angle, area perpendicular to the arrival direction, and time. For a horizontal collection surface $S$, the projected aperture introduces a factor $\cos\theta$. As shown in figure~\ref{fig:TALE_flux}, we adopt the TALE spectrum as the baseline flux model~\cite{ref19}. The number of cosmic rays incident on $S$ during an observation period $[T_1,T_2]$ can then be estimated by
\begin{equation}
\label{eq:Ne0}
N_{\mathrm{ev}}
  = \int_{\mathcal{E}_1}^{\mathcal{E}_2}
    \!\int_{\theta_1}^{\theta_2}
    \!\int_{\phi_1}^{\phi_2}
    \!\int_{T_1}^{T_2}
      J(\mathcal{E})\,S\,\cos\theta\,\sin\theta\,
      d\theta\,d\phi\,d\mathcal{E}\,dt .
\end{equation}
Here, $[\mathcal{E}_1,\mathcal{E}_2]$, $[\theta_1,\theta_2]$, and $[\phi_1,\phi_2]$ denote the energy, zenith-angle, and azimuth-angle ranges, respectively. The measured flux $J(\mathcal{E})$ is shown in figure~\ref{fig:TALE_flux}. In this work the TALE spectrum is used only as the empirical flux input for event-rate integration over the energy range of interest; no interpretation of local spectral structures is required for the FARSim methodology.

For a radio array, the trigger efficiency $\eta(\mathcal{E},\theta,\phi)$ is the fraction of shower cores within a specified fiducial area that satisfy the array trigger at fixed primary energy and arrival direction. The expected number of triggered events is
\begin{equation}
\label{eq:Ne1}
N_{\mathrm{ev}}^{\mathrm{trig}}
  = \int_{\mathcal{E}_1}^{\mathcal{E}_2}
    \!\int_{\theta_1}^{\theta_2}
    \!\int_{\phi_1}^{\phi_2}
    \!\int_{T_1}^{T_2}
      \eta(\mathcal{E},\theta,\phi)\,
      J(\mathcal{E})\,S\,\cos\theta\,\sin\theta\,
      d\theta\,d\phi\,d\mathcal{E}\,dt .
\end{equation}
The local trigger may be defined by a field, signal-to-noise, filtered-voltage, or waveform-level criterion. The array trigger additionally requires at least $N_{\mathrm{trig}}$ stations to satisfy that local condition. If $M_C(\mathcal{E},\theta,\phi)$ core positions are tested in a parameter bin and $M_T(\mathcal{E},\theta,\phi)$ satisfy the array trigger, then
\begin{equation}
\label{eq:eta}
\eta(\mathcal{E},\theta,\phi)
  = \frac{M_T(\mathcal{E},\theta,\phi)}{M_C(\mathcal{E},\theta,\phi)} .
\end{equation}

A conventional calculation requires microscopic radio simulations across energy, zenith angle, azimuth, and independent shower realizations. Core positions and candidate layouts add detector-level evaluations of each resulting footprint. The expensive microscopic stage therefore contains
\begin{equation}
\label{eq:Nsim}
N_{\mathrm{full}}
  = N_E N_\theta N_\phi N_R
\end{equation}
full radio simulations, where $N_E$, $N_\theta$, and $N_\phi$ are the sampled parameter counts and $N_R$ is the number of independent shower realizations. If one full simulation requires CPU time $t_{\mathrm{sim}}$ and $N_C$ cores are used efficiently, the corresponding wall-clock time is approximately
\begin{equation}
\label{eq:Tsim}
T_{\mathrm{wall}}^{\mathrm{full}}
  \simeq \frac{N_{\mathrm{full}}\,t_{\mathrm{sim}}}{N_C}.
\end{equation}
The runtime records of the reference simulation set illustrate this scale. For 20 showers at $10^{17.5}\,\mathrm{eV}$, $\theta=60^\circ$, and $\phi=45^\circ$, comprising ten iron and ten proton realizations with 176 observers and $0.5\,\mathrm{ns}$ sampling, the mean ZHAireS CPU time was $3.78\pm0.38\,\mathrm{h}$. The simulations used a relative thinning level of $10^{-5}$ and a maximum statistical weight factor of 0.06. The zenith-dependent runtime and corresponding FARSim query cost are quantified later in the computational-reduction subsection. Core shifts and alternative layouts need not require new microscopic showers if the radio footprint is reusable, but they still multiply the subsequent field-interpolation and trigger evaluations. FARSim reduces both costs by spanning the shower-parameter grid with a compact reference library and by evaluating core offsets through translated trigger regions.

\section{FARSim Method}
\label{sec:fast_eval}

As discussed in section~\ref{sec:trigger_estimation}, the expensive microscopic stage scales with eq.~\eqref{eq:Nsim}, while core offsets and layouts multiply the subsequent detector-level evaluations. FARSim reduces both costs by reusing a limited set of full radio simulations to construct ground-plane footprints and by evaluating many core offsets and candidate layouts with a common trigger model.

Reference ZHAireS simulations first sample the electric field on star-shaped grids in the shower plane. FARSim decomposes these fields into geomagnetic and charge-excess components, selects the corresponding zenith template, applies geomagnetic-angle and energy scaling, and projects the footprint to the ground. A threshold contour is then translated over sampled core positions relative to the candidate layout. Station multiplicities yield the trigger efficiency and, after flux integration, the screening event rate. This construction supports relative layout comparisons without treating the surrogate as a full detector-response simulation.

\subsection{Coordinate Definition and Simulation Framework}

This work employs ZHAireS to simulate the radio-frequency electric field emitted by EAS at each observation point. Let $\hat{\mathbf{v}}$ denote the unit vector along the shower-propagation direction and $\mathbf{B}$ the local geomagnetic-field vector at the observation site. We denote the global Cartesian coordinates by $(X,Y,Z)$ and the shower-based coordinates by $(x',y',z')$.

The normalized right-handed basis relating the two coordinate systems is
\nopagebreak[4]
\begin{equation*}
\hat{\mathbf e}_{x'}=
\frac{\hat{\mathbf v}\times\mathbf B}{|\hat{\mathbf v}\times\mathbf B|},
\qquad
\hat{\mathbf e}_{y'}=\hat{\mathbf v}\times\hat{\mathbf e}_{x'},
\qquad
\hat{\mathbf e}_{z'}=\hat{\mathbf v}.
\end{equation*}
The cylindrical coordinates $(r',\phi',z')$ share the same origin located at $X_{\max}$, the depth of the shower maximum. Here $r'$ denotes the lateral shower-plane radius. The symbol $\rho$ is used below only for wavefront-curvature radii in the geometrical projection formula.

In these shower-based coordinates, the simulation evaluates the field distribution at sample points forming a star-shaped pattern in the $(x',y')$ plane. The complete two-dimensional field map is obtained through shower-plane interpolation. Throughout this work, the term \emph{signal} refers to the simulated electromagnetic field at observation points corresponding to station locations in the array.

\subsection{Reference Shower Library}
\label{sec:reference_library}

The underlying simulation production contains 920 ZHAireS showers, comprising 460 proton and 460 iron primaries. All showers have a primary energy of $10^{17.5}\,\mathrm{eV}$ and a magnetic azimuth of $45^\circ$. The production spans zenith angles from $45^\circ$ to $90^\circ$ in $1^\circ$ steps, with ten independent random realizations for each primary type and zenith angle. The peak-field reference library does not retain these 920 showers as separate runtime entries. The 900 showers between $45^\circ$ and $89^\circ$ are compressed as follows: at each zenith angle and observer index, the ten proton realizations and the ten iron realizations are averaged separately, and the two composition-specific means are then combined with equal weights. The resulting runtime library therefore contains 45 composition-averaged, zenith-indexed templates. Of these, 41 templates cover the validated $45^\circ$--$85^\circ$ range; the $86^\circ$--$89^\circ$ templates are retained only for boundary diagnostics. The exactly horizontal $90^\circ$ simulations are not used to construct the peak-field templates.

The simulations were performed with AIRES 19.04.08 and ZHAireS 1.0.30a using QGSJET-II-04 for high-energy hadronic interactions. A relative thinning level of $10^{-5}$ and a maximum statistical weight factor of 0.06 were applied. The ground observation level was set to $1300\,\mathrm{m}$ above sea level, together with a custom GDAS atmospheric profile. The geomagnetic field was specified explicitly with a strength of $56.568\,\mu\mathrm{T}$, an inclination of $61.907^\circ$, and a declination of $-0.058^\circ$. The kinetic-energy cuts were $3\,\mathrm{MeV}$ for photons and electrons, $10\,\mathrm{MeV}$ for muons, $60\,\mathrm{MeV}$ for mesons, and $120\,\mathrm{MeV}$ for nucleons. Each shower contains 176 virtual observer positions: 160 points arranged along eight shower-plane rays with 20 radial samples per ray, together with 16 additional cross-check positions. The star-grid topology and its normalized radial sampling relative to the event-specific Cherenkov angle are kept fixed, but the physical spacing is not. The observer coordinates are regenerated for each shower from its $X_{\max}$ distance and atmospheric refractivity and are projected conically onto the observation plane. Consequently, neither the shower-plane radial spacing nor the ground-plane spacing remains constant in metres across the zenith-angle range. The radio traces were sampled at $0.5\,\mathrm{ns}$. The main library settings are summarized in table~\ref{tab:reference_library}.

Band-limited electric-field traces are processed offline by multiplying each Cartesian component's Fourier spectrum by a real, even frequency window and applying the inverse FFT. This is zero-phase spectral filtering, rather than a fixed-order IIR or truncated-kernel FIR design, so a conventional filter order does not apply. The original reference-trace preprocessing uses a rectangular $50$--$200\,\mathrm{MHz}$ mask. Where cosine-tapered filtering is specified below, the same passband is retained and each edge is extended by a half-cosine transition sampled on the native FFT bins. Its nominal width is $8\,\mathrm{MHz}$, represented by $N_{\mathrm{tr}}=\max[1,\operatorname{round}(8\,\mathrm{MHz}/\Delta f)]$ bins, where $\Delta f$ is the trace's frequency spacing; the response is zero beyond the transitions. Comparisons explicitly identified as raw or unfiltered do not use either window.

\begin{table}[!t]
\centering
\caption{Configuration of the ZHAireS production and the compact peak-field library used by FARSim.}
\label{tab:reference_library}
\begin{tabularx}{\textwidth}{@{}>{\raggedright\arraybackslash}p{0.34\textwidth}>{\raggedright\arraybackslash}X@{}}
\toprule
Quantity & Setting \\
\midrule
Underlying simulation production & 920 showers: 460 proton and 460 iron \\
Peak-field template construction & 900 showers at $45^\circ$--$89^\circ$: 450 proton and 450 iron \\
Runtime peak-field library & 45 composition-averaged zenith templates; 41 in the validated $45^\circ$--$85^\circ$ range \\
Primary energy and azimuth & $10^{17.5}\,\mathrm{eV}$; magnetic azimuth $45^\circ$ \\
Zenith angles & $45^\circ$--$90^\circ$ in $1^\circ$ steps \\
Random realizations & 10 per primary type and zenith angle before offline averaging \\
Template composition & Equal-weight mean of the proton and iron averages \\
Simulation software & AIRES 19.04.08; ZHAireS 1.0.30a \\
Hadronic model & QGSJET-II-04 \\
Thinning & Relative level $10^{-5}$; maximum weight factor 0.06 \\
Atmosphere and observation level & Custom GDAS profile; $1300\,\mathrm{m}$ above sea level \\
Geomagnetic field & $56.568\,\mu\mathrm{T}$; inclination $61.907^\circ$; declination $-0.058^\circ$ \\
Particle energy cuts & $3\,\mathrm{MeV}$ ($\gamma,e^\pm$); $10\,\mathrm{MeV}$ ($\mu^\pm$); $60\,\mathrm{MeV}$ (mesons); $120\,\mathrm{MeV}$ (nucleons) \\
Observer sampling & Eight rays with 20 Cherenkov-scaled radial points per ray, plus 16 cross-check points; physical spacing varies by shower \\
Time sampling & $0.5\,\mathrm{ns}$ \\
\bottomrule
\end{tabularx}
\end{table}
\FloatBarrier

\subsection{Field Reconstruction and Radiation Mechanism}

FARSim reconstructs the signal in the \emph{shower plane}, which is perpendicular to the shower axis and exposes the approximate symmetries of the radio footprint. In the plane-wave approximation, the \emph{energy fluence} is the time-integrated Poynting flux,
\begin{equation*}
f=\epsilon_0 c\int_{t_1}^{t_2}|\vec{E}(t)|^2\,dt,
\end{equation*}
where $\epsilon_0$ is the vacuum permittivity, $c$ is the speed of light, and $[t_1,t_2]$ is a window containing the band-limited radio pulse; $f$ has units of $\mathrm{J\,m^{-2}}$. The quantities $E_{\mathrm{geo}}$ and $E_{\mathrm{ce}}$ used in the present FARSim analysis are peak electric-field amplitudes, not energy fluences. For inclined showers, \emph{early--late effects} introduce asymmetries in the ground-plane radio footprint because stations on the early and late sides sample the emission at different geometrical distances and atmospheric depths; standard procedures correct these effects before fitting rotationally symmetric lateral distributions~\cite{Huege2019}.

To further reduce the computational cost of radio simulations, a four-arm reconstruction procedure is adopted. The reconstruction is based on the two principal radio-emission mechanisms in an EAS: the geomagnetic radiation induced by the Lorentz force, represented by $E_{\mathrm{geo}}(r')$, and the charge-excess emission, represented by $E_{\mathrm{ce}}(r')$. The total electric field $\vec{E}_T(r',\phi')$ is obtained from the vector superposition of these two components~\cite{refGeoAskaryanModel}:
\begin{equation}
\label{eq:total_field}
\vec{E}_T(r',\phi') = E_{\mathrm{ce}}(r')\,[\cos(\phi')\,\hat{\mathbf e}_{x'} + \sin(\phi')\,\hat{\mathbf e}_{y'}] + E_{\mathrm{geo}}(r')\,\hat{\mathbf e}_{x'} .
\end{equation}
The four-arm decomposition is most transparent when the radio-emission pattern is close to azimuthal symmetry, but in FARSim it is used as a reduced-order approximation rather than as an exact symmetry assumption. In particular, eq.~\eqref{eq:total_field} does not introduce a separate term for the clover-leaf, geo-synchrotron-like polarization component reported for very inclined showers in strong geomagnetic fields~\cite{refChiche}. That component can locally reach approximately 10\% of the total field amplitude in representative very inclined configurations, although its magnitude depends on atmospheric density, observing frequency, geomagnetic-field strength, and shower geometry.

Its practical impact under the present configuration was checked in a complementary component audit of full ZHAireS traces generated for the same site and filtered to $50$--$200\,\mathrm{MHz}$. At stations with a peak amplitude above 2\% of the event maximum, the net fraction outside the dominant $\hat{\mathbf e}_{x'}\parallel\hat{\mathbf v}\times\mathbf B$ direction, defined as $\sqrt{E_{y'}^2+E_{z'}^2}/|\vec{E}|$ at the vector-envelope peak, averaged 0.021 over 40690 stations from 1106 showers in the $60^\circ$--$85^\circ$ interval. Averaging first within each shower and then over showers gave 0.028. This net orthogonal fraction contains charge-excess emission, any third polarization component, and longitudinal residuals, and interference prevents it from being interpreted as an isolated measurement of the third component. It nevertheless shows that the trigger-relevant field is dominated by the $\hat{\mathbf v}\times\mathbf B$ projection for this site, band, and selected zenith range. Moreover, because the validation below uses full microscopic ZHAireS simulations as its reference, any unmodeled third-component contribution is retained in the benchmark footprints and therefore contributes to the reported reconstruction residuals rather than being absent from the accuracy test. The resulting residual morphology is treated as a model systematic. The applicability of the reduced-order model is assessed empirically at the trigger-observable level over the $45^\circ$--$85^\circ$ working range; simulations at larger zenith angles are used only to diagnose the upper boundary and do not establish validity there.

For inclined showers, the center of the radio ring need not coincide with the geometrical shower-axis intersection on the observation plane~\cite{refRefractiveDisplacement}. The reference templates are therefore radio-core recentered before interpolation. At each zenith angle, the peak positions of $|E_{\mathrm{geo}}|$ along the eight star-grid arms are fitted with a common circle in shower-plane coordinates. The observer coordinates and the effective $X_{\max}$ source position are translated together by the fitted center, preserving their relative geometry while reducing the arm-to-arm spread of the ring position. This operation is a refractive-coordinate correction, not an independent measurement of a particle-core displacement. Tests above $85^\circ$ show that translation alone does not remove the remaining peak-amplitude and footprint-shape differences, which motivates the upper limit adopted here.

\noindent\begin{minipage}{\textwidth}
\setlength{\parindent}{1.5em}
\indent Radial symmetry remains an approximation even after this recentering. CoREAS studies have shown that atmospheric refraction can displace the radio symmetry center by approximately $1.5\,\mathrm{km}$ on the ground for showers near $85^\circ$~\cite{refRefractiveDisplacement}. More recent analytical and ZHAireS work shows that altitude-dependent refractivity and Earth curvature can also produce different Cherenkov angles on the early and late sides, with the difference increasing toward the horizon~\cite{refCherenkovCone}. The present recentering corrects the leading translation of the averaged reference footprint but does not model these residual two-dimensional, arm-dependent, or event-dependent distortions. Their net impact is retained in the residuals obtained from validation against full ZHAireS simulations and contributes to the model uncertainty near the upper boundary of the working range.
\end{minipage}

Here $\phi'$ denotes the local azimuth in the shower plane and is distinct from the shower arrival azimuth $\phi$. The direction $\phi'=0^\circ$ is defined along $+\hat{\mathbf e}_{x'}\parallel\hat{\mathbf v}\times\mathbf B$, and $\phi'$ increases from $+\hat{\mathbf e}_{x'}$ toward $+\hat{\mathbf e}_{y'}$, following the right-hand rule about $+\hat{\mathbf e}_{z'}=\hat{\mathbf v}$. By sampling four azimuthal directions at $\phi'_i=90^\circ i$ $(i=0,\ldots,3)$, the components can be decoupled as
\begin{equation}
\label{eq:field_decomposition}
\begin{cases}
E_{\mathrm{geo}}(r') =
\dfrac{1}{4}
\displaystyle\sum_{i=0}^{3}
\vec{E}_T^{\,\phi'_i}(r')\cdot\hat{\mathbf e}_{x'}, \\[6pt]
E_{\mathrm{ce}}(r') =
\dfrac{1}{2}
\left[
\vec{E}_T^{\,\phi'_1}(r') -
\vec{E}_T^{\,\phi'_3}(r')
\right]\cdot\hat{\mathbf e}_{y'} .
\end{cases}
\end{equation}
The signs in eq.~\eqref{eq:field_decomposition} follow directly from eq.~\eqref{eq:total_field}: the four $x'$ projections sum to $4E_{\mathrm{geo}}$, while the difference between the $\phi'=90^\circ$ and $\phi'=270^\circ$ $y'$ projections gives $2E_{\mathrm{ce}}$. The same convention is used in the implementation after applying the common trace-sign convention to all three Cartesian field components. Thus the decomposition preserves the physical vector superposition before any scalar trigger quantity is extracted.

\begin{figure}[!t]
  \centering
  \includegraphics[width=0.95\textwidth]{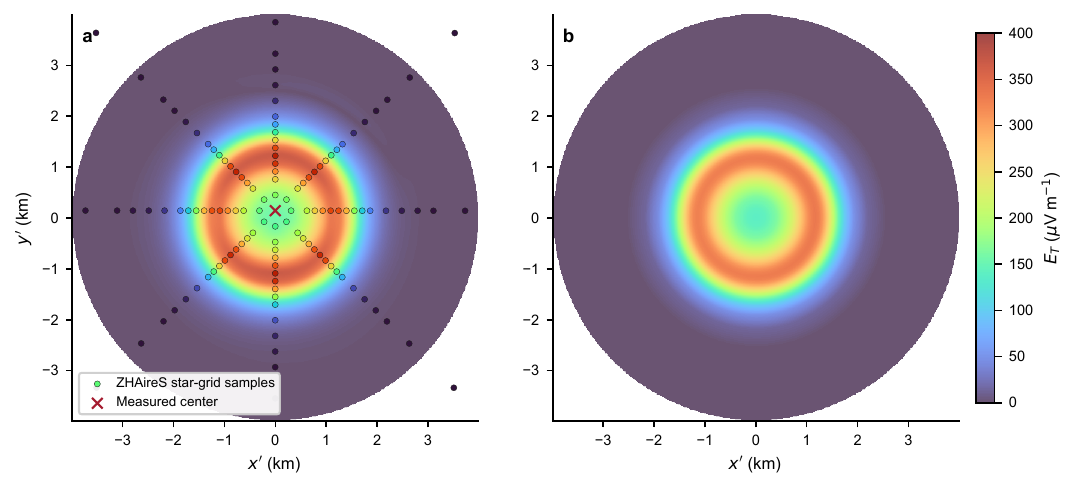}
  \caption{Shower-plane interpolation and FARSim comparison for an iron shower generated with QGSJET-II-04 at $\mathcal{E}=1\,\mathrm{EeV}$, $\theta=84.2^\circ$, magnetic azimuth $\phi=0^\circ$, ZHAireS random seed 0.014439517, and a ground altitude of $1.3\,\mathrm{km}$. The same shower is used in the left column of figure~\ref{fig:egeo-compare}. The raw traces use the cosine-tapered $50$--$200\,\mathrm{MHz}$ FFT filter described in section~\ref{sec:reference_library} before the vector-peak field is extracted. Panel (a) uses all 160 star-grid samples and the central sample; the colored circles show their directly simulated total-field amplitudes, the red cross marks the central sample, and the underlying semitransparent map is obtained by the component-wise PCHIP--makima interpolation described in the text. Panel (b) shows the FARSim prediction constructed from the composition-averaged reference library using the nearest zenith template ($84^\circ$), together with energy and geomagnetic-angle scaling.}
  \label{fig:field_reconstruction}
\end{figure}

Figure~\ref{fig:field_reconstruction} combines the construction of the continuous ZHAireS reference footprint with the validation of the FARSim reconstruction. In panel (a), the colored samples are the directly simulated total-field amplitudes at the discrete ZHAireS observer positions, while the underlying semitransparent continuous map is obtained by interpolating the three shower-plane electric-field components separately. A shape-preserving piecewise-cubic Hermite interpolant (PCHIP) is first applied along each radial arm, with the measured central field used at $r'=0$. The resulting arm profiles are extended periodically in $\phi'$ and interpolated between arms using the modified Akima method (makima). The total field amplitude is then calculated from the three interpolated components and compared with the FARSim reconstruction in panel (b). No zero-valued padding or radial extrapolation is introduced; the comparison is restricted to the radial range sampled by every arm.

Throughout the validation, the Pearson coefficient between matched ZHAireS reference values $Z_i$ and FARSim predictions $F_i$ is defined as
\begin{equation}
\label{eq:pearson_coefficient}
r_{\mathrm P}=
\frac{\sum_i (Z_i-\overline{Z})(F_i-\overline{F})}
{\sqrt{\sum_i (Z_i-\overline{Z})^2\sum_i(F_i-\overline{F})^2}}.
\end{equation}
It measures linear agreement of the matched distributions and is therefore used together with RMSE, which retains the absolute amplitude difference.

\begin{figure}[!t]
  \centering
  \includegraphics[width=0.95\textwidth]{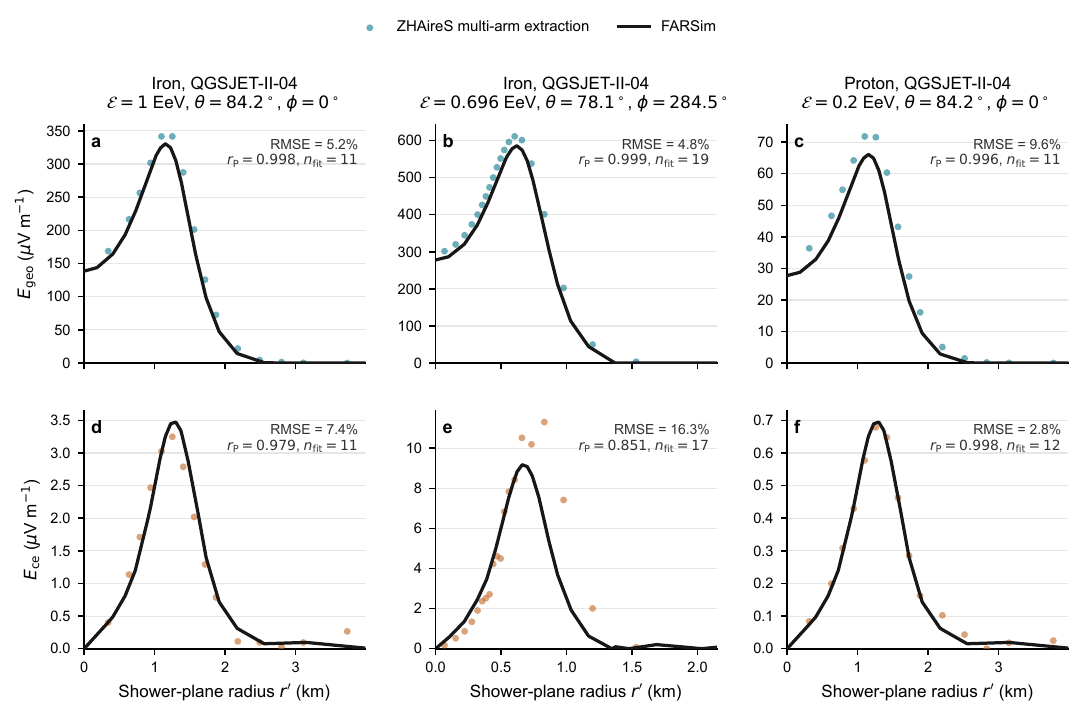}
  \caption{Direct component-level comparisons between full ZHAireS simulations and FARSim for three representative showers. The top row shows $E_{\mathrm{geo}}$ and the bottom row shows $E_{\mathrm{ce}}$. The left column uses the same $1\,\mathrm{EeV}$ iron shower as figure~\ref{fig:field_reconstruction}. The ZHAireS points are the ring-wise multi-arm estimates from eq.~\eqref{eq:validation_multiarm_components}, based on the polarization decomposition of ref.~\cite{refChichePolarisation}; the black curves are the FARSim predictions. The annotated RMSE and Pearson coefficient $r_{\mathrm{P}}$ are evaluated for ring estimates above 5\% of the corresponding ZHAireS maximum and within the FARSim radial support.}
  \label{fig:egeo-compare}
\end{figure}

For the event shown in figure~\ref{fig:field_reconstruction}, the comparison is not only qualitative. The quantitative comparison uses the conservative common circular support $r'\leq3.989\,\mathrm{km}$, set by the shortest of the eight simulated arms and the radial support of the selected FARSim template. Within this domain, the Pearson correlation coefficient between the interpolated ZHAireS field and the FARSim reconstruction is $r_{\mathrm{P}}=0.9981$. The root-mean-square difference is $13.88\,\mu\mathrm{V/m}$, or 3.77\% of the peak reference field, using the same peak-normalized RMSE convention as figure~\ref{fig:egeo-compare}. The ring location and broad footprint morphology are recovered, while the remaining peak-amplitude and arm-dependent deviations illustrate shower-to-shower fluctuations and two-dimensional asymmetries that are absent from the ten-seed-averaged radial template.

\begin{table}[!t]
\centering
\caption{Area-weighted agreement of the threshold regions extracted from the ZHAireS and FARSim maps in figure~\ref{fig:field_reconstruction}. The area ratio is $A_{\mathrm{FARSim}}/A_{\mathrm{ZHAireS}}$; precision and recall treat the ZHAireS threshold region as the reference.}
\label{tab:field_threshold_agreement}
\begin{tabular}{@{}ccccc@{}}
\toprule
Threshold ($\mu\mathrm{V/m}$) & Area ratio & IoU & Precision & Recall \\
\midrule
25 & 0.962 & 0.962 & 1.000 & 0.962 \\
50 & 0.956 & 0.956 & 1.000 & 0.956 \\
75 & 0.975 & 0.975 & 1.000 & 0.975 \\
\bottomrule
\end{tabular}
\end{table}
\FloatBarrier

Although no zero-valued padding enters this comparison, the common low-amplitude outer region can still contribute strongly to a whole-map correlation coefficient. We therefore apply absolute field thresholds of 25, 50, and $75\,\mu\mathrm{V/m}$, all within the dynamic range of both maps, and compare the resulting regions in table~\ref{tab:field_threshold_agreement}. Each polar-grid element is weighted by the radial Jacobian $r'\,\mathrm{d}r'\,\mathrm{d}\phi'$, so the reported quantities describe physical footprint area rather than the number of interpolation nodes. Across the three thresholds, the FARSim-to-ZHAireS area ratio and IoU both range from 0.956 to 0.975; precision is 1.000 and recall is at least 0.956. Thus, the close field-level agreement is also retained after thresholding over this range. These are footprint-level threshold-region metrics, not array-level trigger efficiencies, which additionally depend on the station layout, shower-core distribution, multiplicity requirement, detector response, and noise.

As an additional field-level comparison, figure~\ref{fig:egeo-compare} shows representative $E_{\mathrm{geo}}$ and $E_{\mathrm{ce}}$ profiles obtained from FARSim and directly from the full ZHAireS electric-field traces. The left and right columns use the cosine-tapered FFT window, whereas the middle column retains the rectangular $50$--$200\,\mathrm{MHz}$ window used to preprocess that event. Following the two-component polarization decomposition of ref.~\cite{refChichePolarisation}, the early--late-corrected vector-peak field at observer $i$ is projected onto $\hat{\mathbf e}_{x'}\parallel\hat{\mathbf v}\times\mathbf B$ and $\hat{\mathbf e}_{y'}\parallel\hat{\mathbf v}\times(\hat{\mathbf v}\times\mathbf B)$. For each eight-arm ring, the two amplitudes are estimated as
\begin{equation}
\label{eq:validation_multiarm_components}
\widehat E_{\mathrm{ce}}=
\frac{\sum_i E_{y',i}\sin\phi'_i}{\sum_i\sin^2\phi'_i},
\qquad
\widehat E_{\mathrm{geo}}=
\frac{1}{N}\sum_i\left(E_{x',i}-\widehat E_{\mathrm{ce}}\cos\phi'_i\right).
\end{equation}
Thus all six non-axis projections constrain $E_{\mathrm{ce}}$, while all eight arms contribute to $E_{\mathrm{geo}}$. This is a direct peak-field decomposition and does not infer field amplitudes by taking square roots of parameterized energy fluences.

For the common event shown in the left column of figure~\ref{fig:egeo-compare}, the peak-normalized RMSE values are 5.2\% for $E_{\mathrm{geo}}$ and 7.4\% for $E_{\mathrm{ce}}$. These ring-wise component metrics average the eight star-grid arms at each radius, whereas figure~\ref{fig:field_reconstruction} retains the arm-dependent amplitudes in the two-dimensional total-field map. The two diagnostics are therefore complementary: figure~\ref{fig:egeo-compare} tests the reduced radial component profiles, while figure~\ref{fig:field_reconstruction} also exposes single-shower azimuthal fluctuations around those profiles.

The present comparison concerns trigger-oriented peak-field reconstruction. Calibration of the reduced-order profiles for cosmic-ray energy reconstruction is outside the scope of this study.

\subsection{Unified Scaling and Projection Framework}

To reconstruct radio signals over the parameter space relevant to layout studies, FARSim applies three transformations to a reference simulation: geomagnetic-angle scaling, energy scaling, and ground projection. Together, they define a reduced-order forward model for comparative layout studies rather than a complete detector-response simulation.

First, FARSim does not apply an analytical zenith-angle scaling. The evaluation grid uses integer zenith angles and directly selects the corresponding composition-averaged template from the one-degree reference grid described above; it does not scale a single footprint across zenith angles. Each template therefore carries its own zenith-dependent field amplitude, radial structure, and projection geometry. At this fixed zenith angle, the geomagnetic component is then scaled from the reference azimuth $\phi_0$ to the target azimuth $\phi$ according to the corresponding geomagnetic angles,
\begin{equation}
\label{eq:azimuth_scaling}
E_{\mathrm{geo}}^{\alpha_\phi}(r') = E_{\mathrm{geo}}^{\alpha_{\phi_0}}(r') \cdot \frac{\sin(\alpha_\phi)}{\sin(\alpha_{\phi_0})}.
\end{equation}
The dominant zenith-angle dependence is therefore inherited directly from the zenith-specific reference template, whereas eq.~\eqref{eq:azimuth_scaling} assumes that the remaining geomagnetic-angle dependence is separable from it. In the coherent-emission regime at fixed magnetic-field strength, the geomagnetic field amplitude is expected to scale approximately with $\sin\alpha$. A linear dependence on $|\mathbf{v}\times\mathbf{B}|$ is supported by CODALEMA measurements and the same $\sin\alpha_{\mathrm{target}}/\sin\alpha_{\mathrm{reference}}$ field-amplitude correction is used in Radio Morphing, where it has been tested against ZHAireS simulations for inclined-shower configurations with multiple arrival directions~\cite{ref11,ref26,refRadioMorphing2026}. Complementary radiation-energy studies find an approximately $\sin^2\alpha$ dependence for the geomagnetic contribution, as expected for an observable quadratic in the electric field~\cite{refGeoAskaryanModel}.

\begin{figure}[H]
  \centering
  \includegraphics[width=0.95\textwidth]{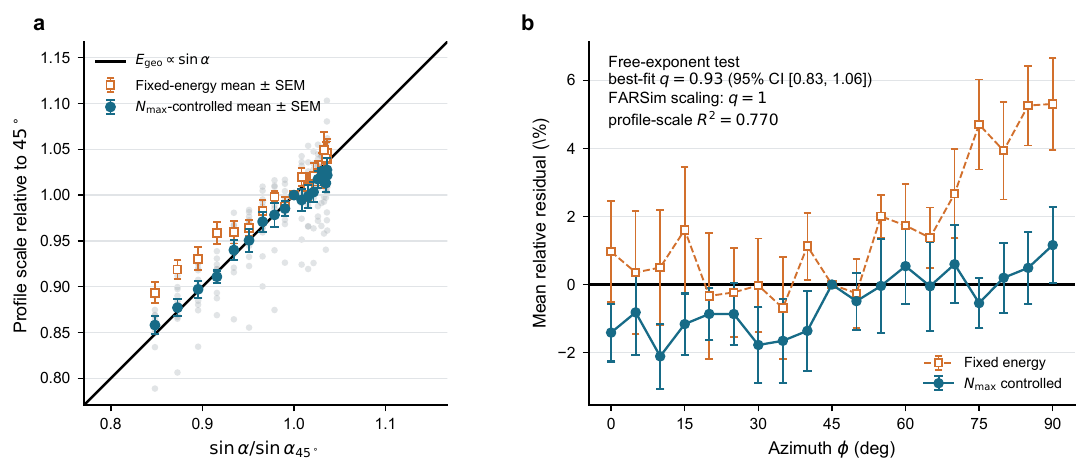}
  \caption{Direct validation of the geomagnetic-angle scaling in eq.~\eqref{eq:azimuth_scaling} using 190 ZHAireS proton simulations at $1.0233\times10^{16}\,\mathrm{eV}$ and $\theta=45^\circ$, grouped into ten matched random-seed sets. Panel (a) compares the profile scale relative to the same seed at $\phi=45^\circ$ with the predicted $\sin\alpha$ ratio. Orange open squares show the fixed-energy means; blue circles show the diagnostic result after controlling for shower-size fluctuations with $N_{\max}$; gray points show the individual controlled values. Panel (b) gives the mean residual relative to the $\sin\alpha$ prediction as a function of azimuth. Error bars denote the standard error of the mean over ten seed groups. The $N_{\max}$ control is used only for this validation and is not part of the FARSim scaling equation. This dedicated simulation set contains only $\theta=45^\circ$ showers.}
  \label{fig:geomagnetic_scaling_validation}
\end{figure}

Figure~\ref{fig:geomagnetic_scaling_validation} presents an additional test of eq.~\eqref{eq:azimuth_scaling} with a dedicated matched simulation set. It contains 190 ZHAireS proton simulations at a fixed primary energy of $1.0233\times10^{16}\,\mathrm{eV}$ and a fixed zenith angle of $45^\circ$, arranged as ten random-seed groups and 19 azimuths from $0^\circ$ to $90^\circ$ in $5^\circ$ steps. Each simulation contains 321 antenna positions. The three-component traces are filtered with a rectangular $50$--$200\,\mathrm{MHz}$ FFT mask, transformed to the shower-plane basis, and sampled at the maximum of the vector field magnitude. At each eight-arm radial ring, $E_{\mathrm{geo}}$ is then estimated with eq.~\eqref{eq:validation_multiarm_components}. For every seed and azimuth, a zero-intercept least-squares factor is fitted between this radial profile and the profile from the same seed at the reference azimuth $45^\circ$; only rings for which both profiles exceed 5\% of their respective maxima are included. The uncorrected fixed-energy profile factors differ from the $\sin\alpha$ prediction by 1.74\% on average and by at most 5.31\% in the azimuth-averaged result.

Because the coherent field amplitude also follows the number of radiating charged particles, the same comparison is repeated after dividing each profile by $N_{\max}$, the number of charged particles at shower maximum reported by ZHAireS. This is a diagnostic control for shower-size fluctuations and is not an additional factor in the FARSim scaling equation. Fitting $E_{\mathrm{geo}}\propto\sin^q\alpha$ to the controlled profiles gives $q=0.93$ with a seed-bootstrap 95\% confidence interval of $[0.83,1.06]$; the mean and maximum relative residuals with respect to the $q=1$ prediction are 0.85\% and 2.11\%, respectively. The result is stable when the radial support threshold is varied from 2\% to 15\% of the profile maximum.

This dedicated test deliberately holds the zenith angle at $45^\circ$ and therefore does not by itself establish the validity of the sine scaling at other zenith angles. Multi-zenith support is supplied by the cited experimental, radiation-energy, and Radio-Morphing studies, while FARSim itself avoids an analytic zenith-angle scaling by selecting the corresponding zenith-specific reference template before applying eq.~\eqref{eq:azimuth_scaling}. Nevertheless, simulations show that stronger magnetic fields and the lower atmospheric densities associated with increasingly inclined showers can enhance transverse particle deflections, causing saturation or coherence loss and making the response depend jointly on zenith angle and $B\sin\alpha$~\cite{refGeoAskaryanModel,refChiche,refRadioMorphing2026}. The Xiaodushan reference simulations use $B=56.568\,\mu\mathrm{T}$; consequently, the sine factor is treated here as a first-order azimuthal correction within the calibrated reference-library configuration, not as a universal scaling law. Residual non-separable effects can become increasingly important toward the upper end of the stated zenith range and require site-specific recalibration or confirmation with full simulations.

Second, because the radio emission is coherent at the considered frequencies, the field amplitude scales approximately linearly with primary energy,
\begin{equation}
\label{eq:energy_scaling}
E(r';\mathcal{E}') =
E(r';\mathcal{E})\frac{\mathcal{E}'}{\mathcal{E}}.
\end{equation}

This energy dependence is supported by early theoretical work and by modern simulations and measurements~\cite{ref26,ref25}. Combining the zenith-specific template selection with the geomagnetic-angle and energy scalings, the shower-plane component amplitudes before ground projection can be summarized as
\begin{equation}
\label{eq:component_scaling_summary}
\begin{aligned}
E_{\mathrm{geo}}(r';\mathcal{E},\theta,\phi)
&=
E_{\mathrm{geo}}^{\mathrm{template}}(r';\theta)
\frac{\mathcal{E}}{\mathcal{E}_{\mathrm{ref}}}
\frac{\sin\alpha(\theta,\phi)}
     {\sin\alpha(\theta,\phi_0)},\\
E_{\mathrm{ce}}(r';\mathcal{E},\theta)
&=
E_{\mathrm{ce}}^{\mathrm{template}}(r';\theta)
\frac{\mathcal{E}}{\mathcal{E}_{\mathrm{ref}}}.
\end{aligned}
\end{equation}
Here $\mathcal{E}_{\mathrm{ref}}$ and $\phi_0$ are the primary energy and arrival azimuth of the reference template, respectively. In the present approximation, no geomagnetic-angle factor is applied to the charge-excess component. Equation~\eqref{eq:component_scaling_summary} summarizes only the component-amplitude scaling in the shower plane; vector recombination follows eq.~\eqref{eq:total_field}, and the subsequent ground projection follows eqs.~\eqref{eq:field_projection} and~\eqref{eq:projection_simplified}.

For each zenith angle, the reference source position used in the geometrical projection is obtained by averaging the $X_{\max}$ positions over the reference-shower realizations at that zenith angle. Consequently, each zenith-specific template has its own mean reference distance, and eq.~\eqref{eq:energy_scaling} changes the field amplitude while retaining this zenith-dependent mean geometry.

Finally, the field is projected from the shower plane to the ground plane. As a first-order geometrical-spreading approximation, the magnitude of the electric field $E_2$ at a point $A'$ on wavefront $\phi_2$ can be estimated from that on wavefront $\phi_1$ by
\begin{equation}
\label{eq:field_projection}
|E_2| = |E_1| \sqrt{\frac{\rho_1 \rho_2}{(\rho_1 + s)(\rho_2 + s)}},
\end{equation}
where $\rho_1$ and $\rho_2$ are the radii of curvature along the two principal directions of the wavefront, and $s$ is the projection distance between $A$ and $A'$. These curvature radii are distinct from the shower-plane lateral coordinate $r'$. For an approximately rotationally symmetric wavefront, this simplifies to
\begin{equation}
\label{eq:projection_simplified}
|E_2| = |E_1| \cdot \frac{\rho}{\rho + s}.
\end{equation}
The quantity $\rho$ is not a fitted parameter. For each shower-plane sample $A$, it is evaluated as the distance from the zenith-specific mean $X_{\max}$ position to $A$, while $\rho+s$ is the distance from the same source position to the corresponding projected ground point $A'$. Thus, $\rho$ is determined separately by the reference geometry at each zenith angle and by the position of each shower-plane sample. Altitude-dependent refractivity is retained in the underlying ZHAireS reference simulations, but this analytical projection does not retrace rays through the refractivity gradient or reconstruct the resulting two-dimensional footprint distortions. Their residual impact is assessed through comparison with full simulations and becomes increasingly relevant toward $85^\circ$~\cite{refRefractiveDisplacement,refCherenkovCone}. This projection maps the reference shower-plane footprint to the ground plane before threshold extraction.

\begin{figure}[!t]
  \centering
  \includegraphics[width=0.98\textwidth]{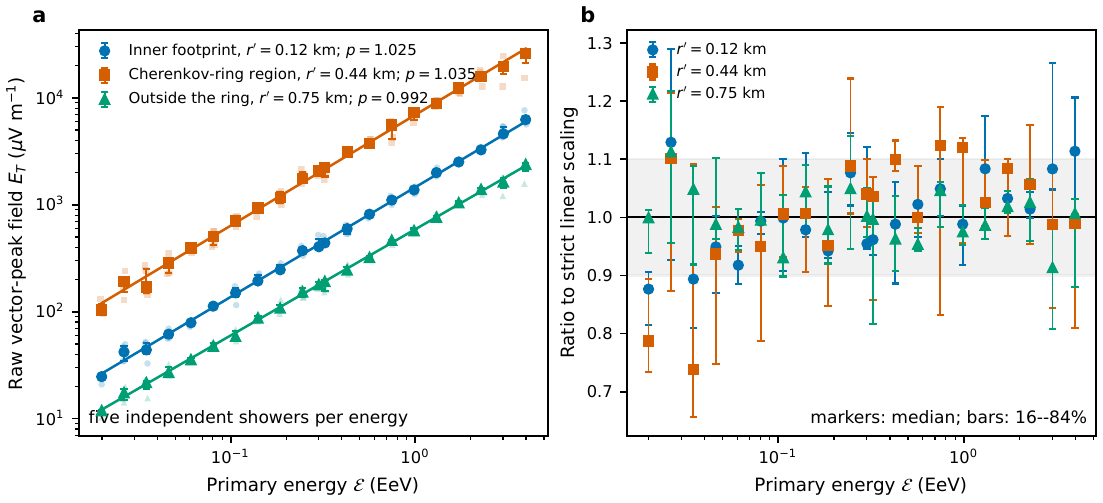}
  \caption{Energy scaling of the raw vector-peak electric field at three fixed shower-plane radii. The dataset contains 105 ZHAireS proton showers generated with SIBYLL~2.3d at fixed $\theta=74.14^\circ$ and $\phi=72.00^\circ$: 21 energies from $10^{16.3}$ to $10^{18.6}\,\mathrm{eV}$ and five independent random seeds per energy. The field is interpolated along the same star-grid arm to the displayed $r'$ values. Faint symbols show individual showers; opaque markers and error bars show the median and 16th--84th percentiles at each energy. Panel (a) gives free power-law fits $E_T\propto\mathcal{E}^{p}$ to the energy medians. Panel (b) shows the ratio $R$ defined in eq.~\eqref{eq:energy_scaling_ratio}; the gray band marks deviations within $\pm10\%$ of exact linear scaling. All 105 showers are retained.}
  \label{fig:energy_relation}
\end{figure}

\begin{figure}[!t]
  \centering
  \includegraphics[width=0.95\textwidth]{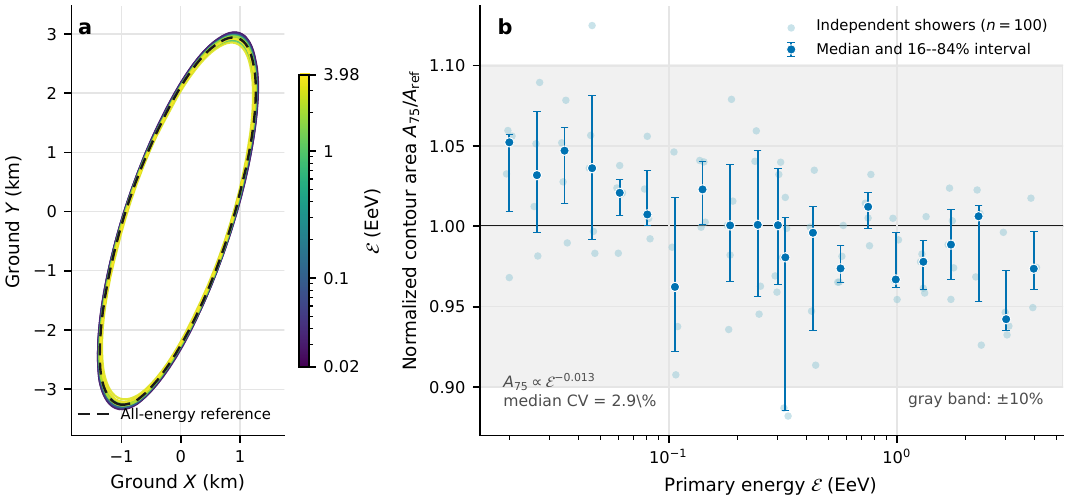}
  \caption{Footprint-level validation of the energy scaling using the same 105-shower scan as figure~\ref{fig:energy_relation}, with the arrival direction fixed at zenith angle $\theta=74.14^\circ$ and magnetic azimuth $\phi=72.00^\circ$. Each raw trace is filtered to 50--200\,MHz, its vector-peak field is normalized to $1\,\mathrm{EeV}$ according to eq.~\eqref{eq:energy_scaling}, and a $75\,\mu\mathrm{V/m}$ ground-plane contour is extracted. Panel (a) shows, for all 21 energies, the contour of the pixel-wise median of the five normalized footprints; color denotes the original primary energy and the black dashed curve is the all-energy median-field reference. Panel (b) shows individual contour areas normalized by the reference area, together with the per-energy median and 16th--84th percentiles. Individual contours are retained only when their maximum radius is at most 98\% of the available radial support; 100 of 105 pass this predefined support requirement, and every energy retains at least three showers.}
  \label{fig:triggerable_region}
\end{figure}

Representative examples supporting the energy scaling are shown in figures~\ref{fig:energy_relation} and~\ref{fig:triggerable_region}. The dedicated fixed-geometry scan contains 105 ZHAireS proton showers generated with SIBYLL~2.3d at the Mazongshan site (ground altitude $2.1\,\mathrm{km}$). It covers 21 primary energies from $10^{16.3}$ to $10^{18.6}\,\mathrm{eV}$, with five independently seeded showers at each energy. The zenith angle is fixed at $\theta=74.14^\circ$ and the magnetic azimuth at $\phi=72.00^\circ$. Thus, primary type, hadronic model, site, and arrival direction are held constant, while energy is varied and the five independent \texttt{RandomSeed} values at each energy sample shower-to-shower fluctuations. This scan isolates the energy dependence under one controlled geometry; it is not, by itself, a test of uniform accuracy over the full primary and directional parameter space.

For figure~\ref{fig:energy_relation}, the raw three-component vector-peak field is linearly interpolated along one common star-grid arm to shower-plane radii $r'=0.12$, 0.44, and $0.75\,\mathrm{km}$, representing the inner footprint, Cherenkov-ring region, and region outside the ring, respectively. Fixing $r'$ avoids comparing different physical observation positions when the generated star-grid coordinates change slightly between showers. At each energy and radius, the five individual-shower values are retained; the plotted opaque marker is their median and the error bar spans the 16th--84th percentiles. No shower is removed from this field-amplitude comparison.

In figure~\ref{fig:energy_relation}(a), each series is fitted with $E_T=C\mathcal{E}^{p}$, where $E_T$ is the maximum Euclidean norm of the three components in the raw ZHAireS electric-field trace. Panel (b) tests the unit-slope scaling through
\begin{equation}
\label{eq:energy_scaling_ratio}
R(r',\mathcal{E}) =
\frac{E_T(r',\mathcal{E})/E_{T,0}(r')}{\mathcal{E}/\mathcal{E}_0},
\end{equation}
where $\mathcal{E}_0=1\,\mathrm{EeV}$ and $E_{T,0}(r')$ is the unit-slope log-space normalization fitted to the energy medians at $\mathcal{E}_0$. Exact linear scaling therefore gives $R=1$. Fits to the 21 energy medians give $p=1.025$, 1.035, and 0.992 at the three radii. The five-shower intervals are wider near the Cherenkov ring because small stochastic changes in shower development shift the ring position across a steep lateral-field gradient and also change its peak amplitude. This behavior is consistent with Radio Morphing, which applies a first-order linear electric-field scaling with primary energy and reports larger residual scaling uncertainty near the Cherenkov ring; a later dedicated Radio-Morphing study further found antenna-level amplitude changes of order 10\% among showers with the same energy, mass, and arrival direction but different random seeds, despite similar LDF shapes~\cite{ref26,refRadioMorphing2026}.

The collapse of the normalized contours in figure~\ref{fig:triggerable_region}(a) and the area statistic in panel (b) provide complementary shape- and area-level tests. The supported individual contours give $A_{75}\propto\mathcal{E}^{-0.013}$, and the coefficient of variation of the 21 per-energy median areas is 2.9\%. The residual energy trend is therefore small compared with the shower-to-shower spread. Together with figure~\ref{fig:energy_relation}, these results support the first-order linear field-amplitude scaling for this fixed primary and arrival direction while retaining the physically larger fluctuations near the Cherenkov ring. Generalization across primary type, zenith angle, and energy is assessed separately below with the independent event sample.

\begin{figure}[!t]
  \centering
  \includegraphics[width=0.95\textwidth]{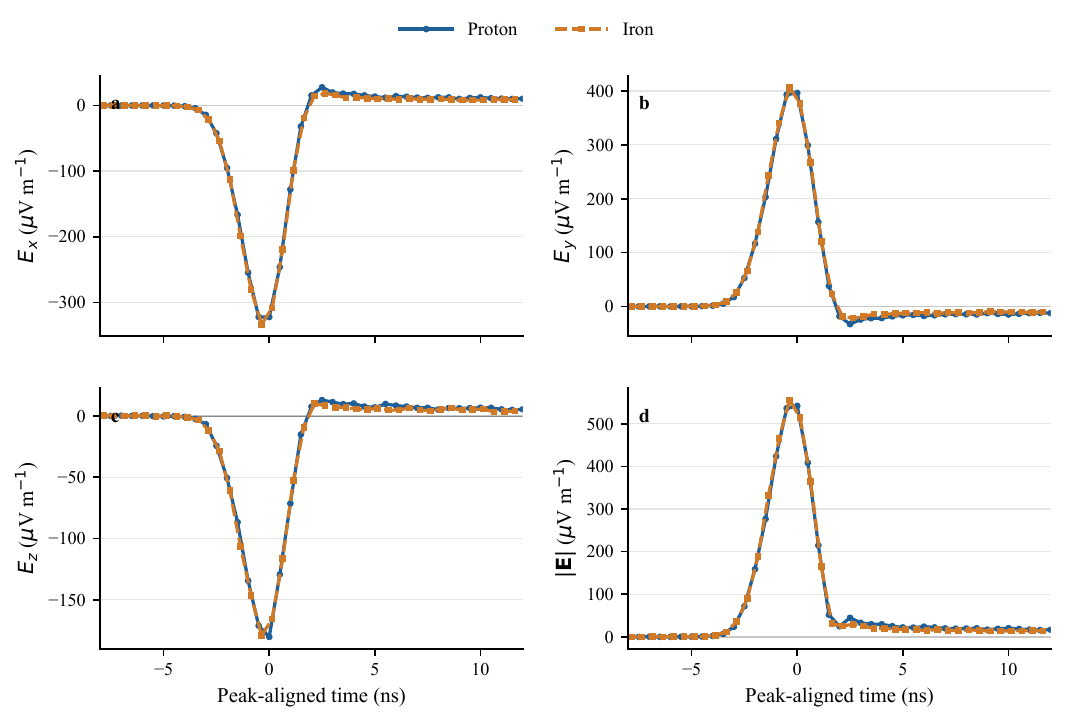}
  \caption{Distance-corrected comparison of raw, unfiltered ZHAireS electric-field pulses at the matched star-grid observer A32 for proton and iron showers generated with QGSJET-II-04 at $\mathcal{E}=10^{17.5}\,\mathrm{eV}$, $\theta=75^\circ$, $\phi=45^\circ$, and random seed 2. The ground observation level is $1.3\,\mathrm{km}$ above sea level. The proton and iron $X_{\max}$-to-observer distances are 42.485 and $47.533\,\mathrm{km}$, respectively. Both traces are corrected to the common distance $R_{\mathrm{ref}}=45.009\,\mathrm{km}$ using $E_{\mathrm{corr}}=E_{\mathrm{obs}}R/R_{\mathrm{ref}}$; no filtering or additional amplitude normalization is applied. The iron trace is shifted by $-0.362\,\mathrm{ns}$ for peak alignment. Panels (a)--(c) show the Cartesian components and panel (d) shows the vector magnitude.}
  \label{fig:time_pulses}
\end{figure}

Figure~\ref{fig:time_pulses} provides a controlled check of the source-distance contribution to a proton--iron waveform comparison. The proton shower has $X_{\max}=729.0\,\mathrm{g\,cm^{-2}}$, whereas the iron shower has $X_{\max}=596.2\,\mathrm{g\,cm^{-2}}$. At observer A32, the corresponding source distances are 42.485 and $47.533\,\mathrm{km}$. Before the first-order $1/R$ correction, the vector-peak amplitudes differ by 9.1\%. After correction to the common reference distance, the peaks are 541.8 and $555.7\,\mu\mathrm{V/m}$, a difference of 2.5\%, while the peak-aligned three-component waveforms have $r_{\mathrm{P}}=0.9992$. Thus, much of the raw amplitude difference in this controlled pair is attributable to the different geometrical distances from $X_{\max}$ to the observer. This single-pair diagnostic is not intended to establish composition independence or to remove all shower-to-shower effects.

Source distance is not neglected in FARSim. For each zenith angle, the reference source position is the mean $X_{\max}$ position of the corresponding reference realizations, and the shower-plane-to-ground projection applies the geometrical-spreading correction in eqs.~\eqref{eq:field_projection} and~\eqref{eq:projection_simplified}. The dominant zenith-dependent mean distance is therefore embedded in the reference template and projection geometry. The present implementation does not, however, apply an additional event-by-event amplitude correction using the true composition-dependent $X_{\max}$ of a queried shower; residual source-distance and shower-development fluctuations remain part of the validation error.

\subsection{Validation of Field Reconstruction, Scaling, and Projection}
\label{sec:field_validation}

The complete forward model, including field decomposition, geomagnetic-angle scaling, energy scaling, and ground projection, is evaluated against full ZHAireS simulations. The event-level analysis tests the lateral distribution of the geomagnetic peak-field amplitude $E_{\mathrm{geo}}$ using the same two-component decomposition, peak-normalized RMSE, and Pearson coefficient $r_{\mathrm{P}}$ as figure~\ref{fig:egeo-compare}. The representative comparison applies the decomposition ring-wise to the star-grid arms, whereas the independent detector-layout events use each station's local shower-plane azimuth.

The validation set contains 1999 ZHAireS showers that are separate from the reference-library events and have randomly sampled directions. Each event uses one seed drawn from ten available values; the seeds are distributed across the sample rather than replicated at every energy and direction. After restricting the analysis to $60^\circ\leq\theta\leq85^\circ$, retaining stations within the radial support for which the reference $E_{\mathrm{geo}}$ exceeds 5\% of the event maximum, and requiring at least eight retained stations, 941 events contribute numerical metrics: 503 iron and 438 proton showers. Their energies span $10^{15}$--$3.999\times10^{18}\,\mathrm{eV}$. For the event-level comparison, both the reference-template traces and the validation traces use the cosine-tapered $50$--$200\,\mathrm{MHz}$ FFT window described in section~\ref{sec:reference_library}. The reference profiles are averaged over ten seeds within each primary type. Template and validation coordinates use the same radio-core convention: reference coordinates use ring-fit recentering, and validation-station coordinates include the event-specific refractive radio-core shift before shower-plane projection.

For each retained event we calculate the peak-normalized relative root-mean-square error
\begin{equation}
\label{eq:event_relative_rmse}
\mathrm{RMSE}_{\mathrm{rel}}=
\frac{\sqrt{N_{\mathrm{fit}}^{-1}\sum_i\left(E_{\mathrm{geo},i}^{\mathrm{FARSim}}-E_{\mathrm{geo},i}^{\mathrm{ZHAireS}}\right)^2}}
{\max_i E_{\mathrm{geo},i}^{\mathrm{ZHAireS}}},
\end{equation}
and the Pearson correlation coefficient $r_{\mathrm{P}}$ over the same matched stations. The former measures the amplitude-weighted profile difference, whereas $r_{\mathrm{P}}$ measures the linear agreement and therefore mainly tests whether the lateral-profile shape is preserved. Because the retained stations are selected relative to each event's own reference peak rather than by an absolute local-trigger cut, these continuous event-level metrics are not tied to the $50\,\mu\mathrm{V/m}$ threshold used in the original validation. The dependence of the reconstructed footprint region on an absolute field threshold is assessed separately at 25, 50, and $75\,\mu\mathrm{V/m}$ in table~\ref{tab:field_threshold_agreement}.

\begin{figure}[!t]
  \centering
  \includegraphics[width=0.95\textwidth]{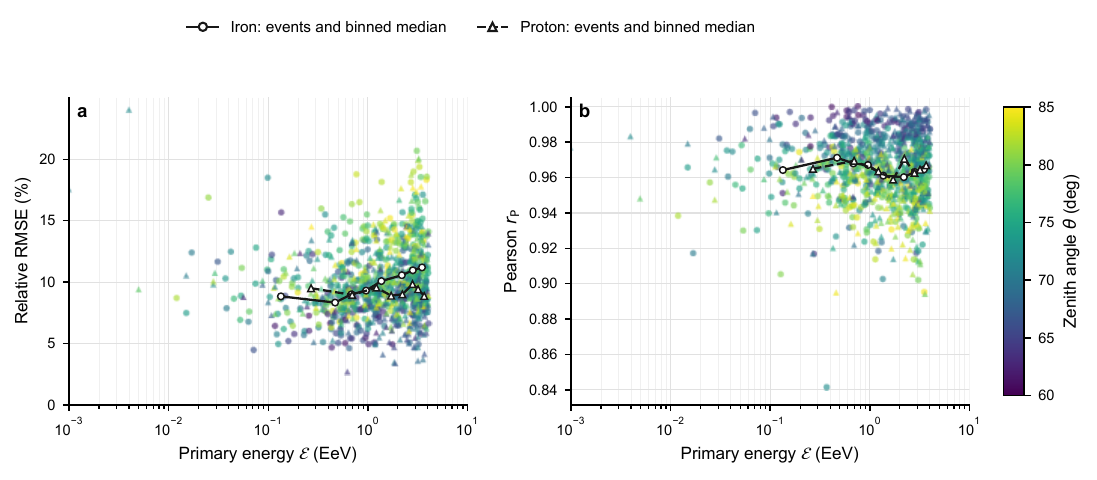}
  \caption{Event-level agreement between the FARSim and ZHAireS $E_{\mathrm{geo}}$ lateral profiles as a function of primary energy. Panel (a) shows the peak-normalized relative RMSE and panel (b) the Pearson correlation coefficient $r_{\mathrm{P}}$. The validation set contains 1999 independent showers that are separate from the reference-library events; 941 events (503 iron and 438 proton) satisfy the predefined common-support and $N_{\mathrm{fit}}\geq8$ requirements and contribute to the plotted metrics. Each event uses one seed drawn from ten available values. Only stations within the common radial support and above 5\% of the event's reference peak are used. Point color denotes zenith angle; circles and triangles denote iron and proton showers, respectively. Black solid and dashed curves show the energy-binned medians for iron and proton events.}
  \label{fig:event_metric_energy}
\end{figure}

Figure~\ref{fig:event_metric_energy} shows both metrics as functions of primary energy. Each point represents one event, and the color encodes its zenith angle. We use one symbol family for correlation coefficients: $r_{\mathrm{P}}$ denotes the Pearson coefficient defined in eq.~\eqref{eq:pearson_coefficient}, while $r_{\mathrm{S}}$ denotes the Spearman rank coefficient, equivalently the Pearson coefficient applied to rank-transformed variables, and tests monotonic trends without assuming linearity. The energy dependence is weak over the full sample: $r_{\mathrm{S}}=0.149$ for energy versus relative RMSE and $r_{\mathrm{S}}=-0.041$ for energy versus $r_{\mathrm{P}}$. The corresponding zenith-angle coefficients are $r_{\mathrm{S}}=0.396$ and $-0.592$, respectively, indicating a stronger zenith-angle dependence. In $5^\circ$ zenith bins, the median relative RMSE increases from 7.8\% at $60^\circ$--$65^\circ$ to 10.6\% at $80^\circ$--$85^\circ$, while the median $r_{\mathrm{P}}$ decreases from 0.985 to 0.950. Thus, zenith angle contributes more strongly to the residual spread than energy over this sample.

\begin{figure}[!t]
  \centering
  \includegraphics[width=0.95\textwidth]{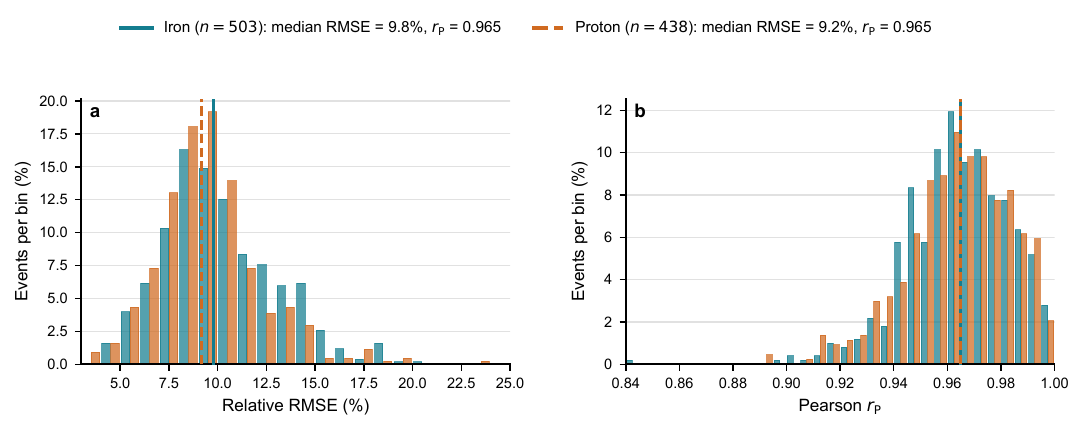}
  \caption{Event-level distributions of (a) the peak-normalized relative RMSE and (b) the Pearson correlation coefficient $r_{\mathrm{P}}$ for the same 941-event sample as figure~\ref{fig:event_metric_energy}. Histograms use common bins and are normalized separately by the number of iron and proton events, so the vertical axis gives the percentage of each primary sample in a bin. Solid and dashed vertical lines mark the corresponding iron and proton medians. The distribution widths include shower-to-shower fluctuations associated with the randomly selected seed as well as residual geometry, interpolation, and template errors.}
  \label{fig:event_metric_distribution}
\end{figure}

The corresponding event-level distributions are shown in figure~\ref{fig:event_metric_distribution}. Iron events have a median relative RMSE of 9.8\%, with a 16th--84th percentile interval of 7.5\%--13.4\%, and a median $r_{\mathrm{P}}$ of 0.965 with an interval of 0.946--0.984. Proton events give 9.2\% (7.2\%--11.8\%) and 0.965 (0.946--0.983), respectively. The percentile intervals contain the central 68\% of events; they describe the observed event-to-event dispersion and are not confidence intervals.

Random-seed selection contributes physically to this dispersion because it changes the stochastic shower development, including $X_{\max}$ and the number of radiating particles. This scale is consistent with the dedicated Radio Morphing study, whose figure~4 compares three showers with identical energy, mass, and arrival direction but different random seeds: the LDF shape changes little, whereas the antenna-level electric-field amplitude varies by roughly 10\%~\cite{refRadioMorphing2026}. The widths in figure~\ref{fig:event_metric_distribution} should therefore be interpreted as seed-inclusive empirical errors. They do not isolate a seed-only uncertainty, because the present sample also spans different energies and directions and includes residual interpolation, finite-support, and template-approximation errors.

\begin{figure}[!t]
  \centering
  \includegraphics[width=0.95\textwidth]{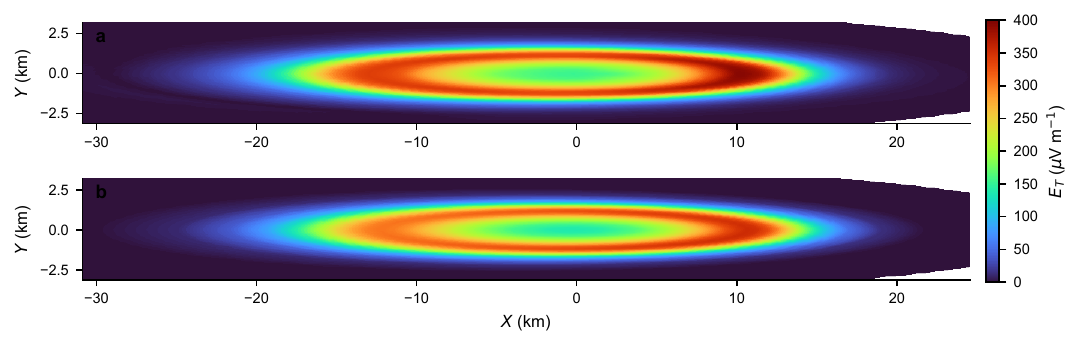}
  \caption{Ground-plane field comparison for the same QGSJET-II-04 iron shower as in figure~\ref{fig:field_reconstruction}: $\mathcal{E}=1\,\mathrm{EeV}$, $\theta=84.2^\circ$, magnetic azimuth $\phi=0^\circ$, ZHAireS random seed 0.014439517, and ground altitude $1.3\,\mathrm{km}$. Panels (a) and (b) are obtained by projecting the interpolated ZHAireS and FARSim shower-plane maps, respectively, to the observation level with the geometrical-spreading correction.}
  \label{fig:field_footprint_comparison}
\end{figure}

\begin{figure}[!t]
  \centering
  \includegraphics[width=0.95\textwidth]{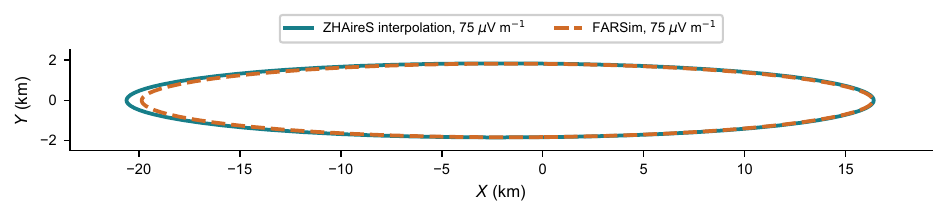}
  \caption{Comparison of the $75\,\mu\mathrm{V/m}$ contours extracted from the interpolation-based ZHAireS footprint and the FARSim reconstruction for a QGSJET-II-04 iron shower with $\mathcal{E}=1\,\mathrm{EeV}$, $\theta=84.2^\circ$, magnetic azimuth $\phi=0^\circ$, ZHAireS random seed 0.014439517, and ground altitude $1.3\,\mathrm{km}$.}
  \label{fig:trigger_region_comparison}
\end{figure}

To complement the field-level validation above, we further examine whether the reconstructed ground-plane footprint preserves the main trigger-relevant spatial features by comparing representative signal distributions and trigger regions. Figure~\ref{fig:field_footprint_comparison} projects the two shower-plane maps in figure~\ref{fig:field_reconstruction} to the observation level, and figure~\ref{fig:trigger_region_comparison} extracts the $75\,\mu\mathrm{V/m}$ contours from those same maps. Together, the field maps and trigger contours test whether the proposed approximation preserves the geometry relevant to multiplicity-based trigger studies.

Together, figures~\ref{fig:egeo-compare}--\ref{fig:trigger_region_comparison} test the reconstruction at the radial-profile, two-dimensional footprint, and threshold-region levels. The resulting metrics support its use for comparative trigger screening within the stated parameter range. They do not constitute an absolute detector-exposure validation, because near-threshold migrations, seed-dependent shower fluctuations, detector noise, and hardware trigger logic can alter the final exposure. The event rates below are therefore reported as screening metrics under explicit trigger abstractions rather than as detector-performance predictions.

\subsection{Trigger Determination and Core Sampling}

Once the ground-plane field has been reconstructed, the trigger condition can be evaluated in a simplified but efficient manner. Here FARSim is intended as a layout-oriented trigger surrogate rather than a full end-to-end detector simulation. Accordingly, the trigger model is formulated in terms of threshold exceedance and station multiplicity under a common set of assumptions, so that different candidate layouts can be compared on a consistent basis. More detailed detector effects, such as site-dependent noise variability, full electronics response, or advanced self-trigger logic, are beyond the scope of the present framework and can be incorporated in future extensions if needed.

Instead of recalculating fields for every shower-core position, FARSim shifts the trigger contour over randomly sampled core locations within an extended array boundary. For each offset, the layout coordinates are tested against the trigger contour, and the event is counted as detected when the number of enclosed stations is not smaller than the multiplicity threshold $N_{\mathrm{trig}}$.

\begin{figure}[!t]
  \centering
  \includegraphics[width=0.95\textwidth]{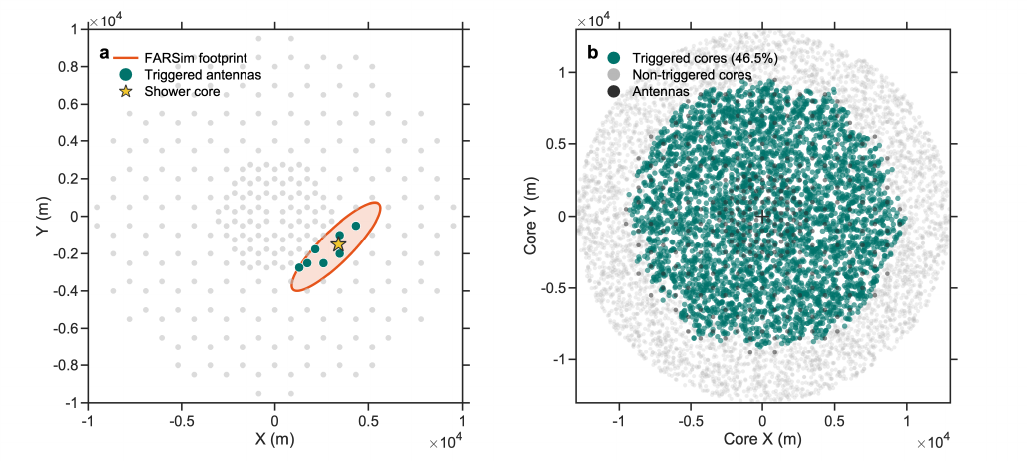}
  \caption{Representative FARSim trigger schematic. Panel (a) shows one triggered core position; the colored footprint denotes the reconstructed ground-plane field distribution, the contour denotes the threshold-defined trigger region, and the highlighted stations satisfy the trigger condition. Panel (b) shows all 8000 core positions sampled uniformly in area, of which 3722 satisfy the five-station trigger condition, corresponding to a trigger efficiency of 46.5\%.}
  \label{fig:trigger_efficiency}
\end{figure}

The evaluation of trigger efficiency is illustrated in figure~\ref{fig:trigger_efficiency}. Panel (a) shows a FARSim trigger region placed on the array layout and the corresponding locally triggered stations for one representative core position. Panel (b) shows all 8000 core positions sampled uniformly in area; 3722 satisfy the five-station trigger condition, corresponding to a trigger efficiency of 46.5\%. Repeating this procedure over the sampled parameter grid yields the trigger efficiency, the average number of triggered stations, and the event-rate estimates used in later array-level studies.

\subsection{Computational Reduction and Fast Event Estimation}

Finally, the number of full ZHAireS simulations is reduced because the expensive radio-footprint generation is performed for a limited set of reference showers and then reused across layouts and core offsets. If $N_Z$ denotes the number of reference ZHAireS showers, $N_S$ the number of stochastic samples evaluated by FARSim, $\Delta T$ the full-simulation time per reference shower, and $N_C$ the number of parallel cores, the wall-clock cost of the full-simulation stage scales approximately as $N_Z\Delta T/N_C$, while the computationally lighter FARSim stage scales with $N_S$.

For numerical rate estimates, eq.~\eqref{eq:Ne1} is evaluated bin by bin. Let $e=\log_{10}(\mathcal{E}/\mathrm{eV})$, and let $\mathcal{E}_{k,1}$ and $\mathcal{E}_{k,2}$ be the lower and upper edges of an energy bin. Experimental cosmic-ray spectra are often tabulated as $F_{\mathrm{GeV}}(E_{\mathrm{GeV}})=E_{\mathrm{GeV}}^3J_{\mathrm{GeV}}(E_{\mathrm{GeV}})$ with energy measured in GeV and differential flux per GeV. All integrations in this work are carried out in eV, using
\begin{equation}
\label{eq:flux_unit_conversion}
J(\mathcal{E})
  =
  \frac{
    F_{\mathrm{GeV}}(\mathcal{E}/10^9)
  }{
    (\mathcal{E}/10^9)^3\,10^9
  },
\end{equation}
where $J(\mathcal{E})$ is the differential flux per eV. For one energy--zenith--azimuth bin, the expected number of triggered events is approximated by
\begin{equation}
\label{eq:event_number_fast}
\Delta N_{k\ell m}
  \simeq
  \Delta T\,S_{\mathrm{throw}}\,
  \overline{J}_k\,\Delta\mathcal{E}_k\,
  \Delta\Omega^{\cos}_{\ell m}\,
  \eta_{k\ell m},
\end{equation}
where $\Delta T=86400\,\mathrm{s}$ for rates quoted per day, $S_{\mathrm{throw}}$ is the area over which shower cores are thrown, $\Delta\mathcal{E}_k=\mathcal{E}_{k,2}-\mathcal{E}_{k,1}$,
\begin{equation}
\overline{J}_k
  = \frac{1}{2}
  \left[
  J(\mathcal{E}_{k,1})
  + J(\mathcal{E}_{k,2})
  \right],
\qquad
\Delta\Omega^{\cos}_{\ell m}
  = \frac{\phi_{m,2}-\phi_{m,1}}{2}
  \left(\sin^2\theta_{\ell,2}-\sin^2\theta_{\ell,1}\right).
\end{equation}
Here $J$ has units of $\mathrm{m^{-2}\,s^{-1}\,sr^{-1}\,eV^{-1}}$, so that eq.~\eqref{eq:event_number_fast} is dimensionless. The total daily rate is obtained by summing $\Delta N_{k\ell m}$ over all sampled bins. In this work, $\eta_{k\ell m}$ is the FARSim trigger efficiency estimated from thrown-core trials within the same bin; it is therefore a screening-level efficiency under the stated trigger abstraction, not an independently validated detector exposure. For the 65-unit array rates quoted below, the shower cores are thrown uniformly in a disk centered on the layout centroid with radius
\begin{equation}
R_{\mathrm{throw}}=\max_i\sqrt{x_i^2+y_i^2}+750\,\mathrm{m}
  =4.602\,\mathrm{km},
\end{equation}
giving $S_{\mathrm{throw}}=\pi R_{\mathrm{throw}}^2=66.54\,\mathrm{km^2}$. This disk is fixed for all energy, zenith-angle, azimuth-angle and trigger-mode bins in the 65-unit array comparison. It is a fiducial core-throw domain for estimating effectively reconstructable events, rather than an extrapolation to arbitrarily distant cores. When the shower core lies far outside the array, the stations sample only a limited portion of the finite-curvature, approximately hyperbolic radio wavefront; such geometries are difficult to constrain robustly in core, direction, and timing reconstruction. They are therefore deliberately excluded from the screening-rate estimate adopted here.

The computational gain depends on the reference-library strategy. A direct bin-by-bin library requires $N_E N_\theta N_\phi$ full radio footprints, with additional realizations when shower fluctuations are sampled. Optimized microscopic workflows can already reuse each footprint across core offsets and layouts through interpolation. FARSim provides two further reductions. First, $N_Z$ reference footprints span the shower-parameter grid, reducing the expensive footprint-generation stage by approximately $N_E N_\theta N_\phi/N_Z$ relative to a direct bin-by-bin library. In the present implementation, the 900 microscopic showers used for peak-field template construction are reduced offline to 45 zenith-indexed runtime templates. Second, random core offsets are evaluated by translating the extracted trigger contour and counting the enclosed stations. This replaces repeated field interpolation at every shifted station position with point-in-region tests.

The ten-realization average used here is a conservative construction choice rather than a runtime requirement. A stochastic reduced-library strategy has a precedent in the updated Radio Morphing implementation, which draws the $X_{\max}$ grammage from a Gaussian distribution parameterized by its mean and root-mean-square variation and represents the approximately 10\% fluctuation in the number of shower particles by a Gaussian perturbation of the effective target energy~\cite{refRadioMorphing2026}. Motivated by this approach, a further FARSim reduction would be possible if a dedicated calibration showed that the correlated seed-to-seed residuals around a baseline profile were adequately described by a conditional Gaussian model, including their zenith- and radius-dependent covariance. Under that assumption, one baseline realization per primary type and zenith angle could be retained and the stochastic fluctuation sampled from the calibrated residual model, reducing the $45^\circ$--$89^\circ$ construction set from 900 to 90 showers while preserving the same 45 zenith-indexed runtime templates. Radio Morphing does not by itself establish Gaussianity of arbitrary station-wise FARSim residuals, and this reduction is not used in the results reported here.

\begin{figure}[!t]
  \centering
  \includegraphics[width=0.95\textwidth]{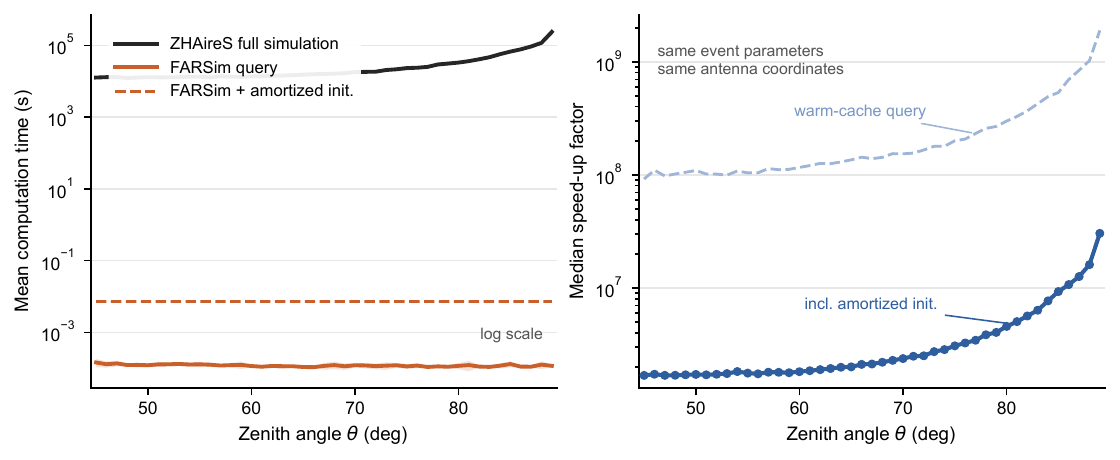}
  \caption{Runtime comparison between full ZHAireS simulations and FARSim field queries as a function of zenith angle. The ZHAireS values are the mean CPU times extracted from the \texttt{.sry} summary files. FARSim is evaluated for the same shower parameters and the same station coordinates as the corresponding ZHAireS event. The dashed FARSim curve includes the one-time template-library initialization cost amortized over the 900 evaluated events.}
  \label{fig:runtime_comparison}
\end{figure}

Figure~\ref{fig:runtime_comparison} compares the ZHAireS CPU time recorded in the simulation summaries with the FARSim time required to evaluate the same event parameters at the same station coordinates. The comparison uses 900 ZHAireS showers with zenith angles from $45^\circ$ to $89^\circ$ and 176 station positions per event. The mean ZHAireS runtime increases from 3.51 h at $45^\circ$ to 65.47 h at $89^\circ$, whereas the FARSim query time remains below one millisecond after template-library initialization. When the one-time initialization cost is amortized over the 900 queries, the median speed-up is $2.2\times10^6$.

The one-time microscopic simulation cost, obtained by summing the recorded CPU times, is approximately $7.7\times10^3$ CPU hours for the full 920-shower production and $7.6\times10^3$ CPU hours for the 900 showers used in peak-field template construction. Restricting construction to the validated $45^\circ$--$85^\circ$ interval would require 820 showers, corresponding to $4.7\times10^3$ recorded CPU hours. For the 900-shower set, ideal event-level parallelism on 40 comparable CPU cores would give about eight days, excluding queueing, I/O, and template post-processing; the actual time depends on hardware and simulation settings. Changing only station positions, thresholds, or the RF-chain response can reuse the shower library within its spatial and physical domain. Changing the site atmosphere, observation altitude, geomagnetic field, or hadronic model requires a new, appropriately configured library and validation, for which the present production gives an order-of-magnitude computational budget.

\subsection{Method Summary and Scope}

FARSim combines field reconstruction, geomagnetic-angle and energy scaling, geometrical projection, and trigger-region evaluation to reduce the number of full radio simulations and repeated field-interpolation calls. The validation above quantifies its field-profile and threshold-region performance within the stated parameter range. The next section extends the peak-field model to time-domain electric-field synthesis and voltage-domain trigger evaluation.

\section{Time-Domain Extension and RF-Chain Trigger Diagnostics}
\label{sec:waveform_rfchain}

The field-level framework supports contour-based trigger estimates using peak electric-field amplitude. Detector-level triggering instead depends on the voltage waveform after antenna response, filtering, amplification, digitization, and noise addition. A peak-field surrogate cannot represent pulse width, spectral content, or threshold-crossing logic. FARSim therefore adds time-domain electric-field synthesis without requiring a new ZHAireS simulation at each trial station.

\subsection{Generation of Time-Domain Electric-Field Traces}

For a shower with primary parameters $(\mathcal{E},\theta,\phi)$ and an observation point $\mathbf{r}$, the original FARSim pipeline provides the three-component peak-field vector
\begin{equation}
\label{eq:farsim_peak_vector}
\mathbf{E}_{\mathrm{pk}}^{\mathrm{F}}(\mathbf{r})
  = \left(E_{x,\mathrm{pk}}^{\mathrm{F}},E_{y,\mathrm{pk}}^{\mathrm{F}},E_{z,\mathrm{pk}}^{\mathrm{F}}\right).
\end{equation}
The time-domain extension combines the three-component peak-field vector with a scalar normalized pulse shape,
\begin{equation}
\label{eq:farsim_waveform_model}
\mathbf{E}^{\mathrm{F}}(t,\mathbf{r})
  = \mathbf{E}_{\mathrm{pk}}^{\mathrm{F}}(\mathbf{r})\,
    s\!\left[t-t_0;\,\mathbf{q}(\mathcal{E},\theta,\phi,\mathbf{r})\right],
\end{equation}
where $s$ is normalized to unit peak and $\mathbf{q}$ denotes the station and shower descriptors available to FARSim. The present descriptors comprise the predicted peak amplitude, shower-plane radius, offset from the empirical Cherenkov-ring angle, distance to $X_{\max}$, zenith angle, arrival and shower-plane azimuth information, and primary type. The geomagnetic-angle dependence enters the waveform through the FARSim peak-field vector rather than as a separate pulse-shape feature.

The pulse-shape model is constructed from raw, unfiltered ZHAireS traces sampled at $0.5\,\mathrm{ns}$. Unlike the compact zenith-indexed peak-field library, this station-level waveform model uses traces drawn from the full production of 920 QGSJET-II-04 proton and iron showers at $10^{17.5}\,\mathrm{eV}$ with zenith angles from $45^\circ$ to $90^\circ$. Complete shower events define the split: 736 events construct the pulse-shape libraries, while 184 independent events containing 2112 retained station traces are reserved for pulse-shape validation. For each station, a baseline estimated from the trace edges is removed. The three-component trace is then projected onto its peak-field direction, aligned at the vector-peak time, resampled on a common $0.5\,\mathrm{ns}$ grid, and normalized to unit peak.

Inference uses non-parametric radial-basis-function-weighted nearest-neighbor interpolation in the standardized descriptor space. The broad template library contains 8380 station traces and uses 18 neighbors. To improve the representation of narrow pulses near the Cherenkov ring, this prediction is blended with a ring-focused library containing 64000 traces from the same 736 training showers and using 10 neighbors. The ring contribution is weighted as $w=0.5\exp[-\tfrac{1}{2}(|\Delta\psi|/0.25^\circ)^2]$, where $\Delta\psi$ is the signed angular offset from the empirical ring. During deployment, the descriptors are calculated from the queried geometry and FARSim output, and no ZHAireS trace at the queried station is accessed. Equation~\eqref{eq:farsim_peak_vector} supplies the absolute amplitude and polarization, while the interpolated template supplies only the normalized waveform morphology. Because the present shape library was constructed at $10^{17.5}\,\mathrm{eV}$, broader energy-dependent pulse-shape calibration remains a subject for future work.

Representative station-level waveform checks are provided in appendix~\ref{app:waveform_rfchain_diagnostics}. In these checks, the generated FARSim traces are compared with raw ZHAireS traces for a $10^{17.5}\,\mathrm{eV}$ iron shower with $\theta=75^\circ$ and $\phi=45^\circ$. The displayed vector peak ratios range from 0.86 to 0.96, and the normalized waveform errors in the plotted windows are about 0.01--0.07, indicating that the template-based extension captures the leading pulse structure in the signal-dominated part of the footprint.

\begin{figure}[!t]
  \centering
  \includegraphics[width=0.95\textwidth]{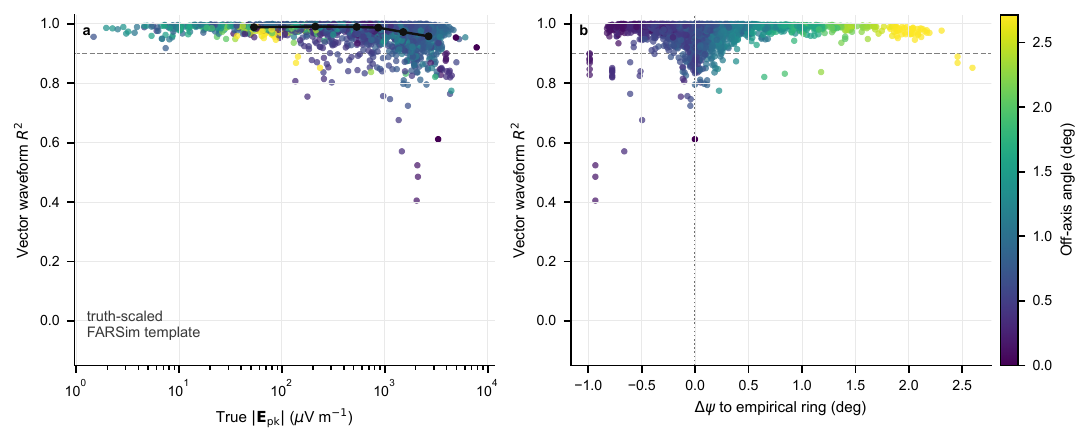}
  \caption{True-amplitude-scaled validation of the FARSim time-domain waveform template. Each point is one station trace from an event held out from pulse-shape model fitting. The FARSim waveform is scaled by the true ZHAireS peak vector before computing $R^2$, so the statistic tests pulse-shape agreement rather than peak-amplitude prediction. Panel (a) shows the three-component vector-waveform $R^2$ as a function of the true peak field; black markers indicate the median in peak-field bins. Panel (b) shows the same $R^2$ as a function of the signed offset from the empirical Cherenkov-ring angle. The dashed horizontal line marks $R^2=0.9$, and the color scale represents the off-axis angle.}
  \label{fig:waveform_r2_truth_scaled}
\end{figure}

To isolate pulse-shape fidelity from peak-field error, the pulse-shape validation scales each FARSim template by the true ZHAireS peak vector. FARSim therefore supplies only the normalized waveform morphology in this test. For the peak-aligned signal window $W=[-80,180]\,\mathrm{ns}$, the three-component waveform score is
\begin{equation}
\label{eq:vector_waveform_r2}
R_{\mathrm{vec}}^2=
1-
\frac{\sum_{j\in W}\sum_{c\in\{x,y,z\}}
\left(E_{c,j}^{\mathrm{ZHAireS}}-E_{c,j}^{\mathrm{FARSim}}\right)^2}
{\sum_{j\in W}\sum_{c\in\{x,y,z\}}
\left(E_{c,j}^{\mathrm{ZHAireS}}-\overline{E}_{c}^{\mathrm{ZHAireS}}\right)^2},
\end{equation}
where $\overline{E}_{c}^{\mathrm{ZHAireS}}$ is the time average of component $c$ within $W$. This direct sample-wise agreement statistic differs from the scale- and offset-invariant Pearson coefficient in eq.~\eqref{eq:pearson_coefficient}: $R_{\mathrm{vec}}^2=1$ denotes exact agreement, $R_{\mathrm{vec}}^2=0$ denotes no improvement over the component-wise temporal mean, and poorer predictions may yield negative values. No additional time-shift optimization is applied to the reported score. As shown in figure~\ref{fig:waveform_r2_truth_scaled}, the geometry-dependent template used here gives a median vector-waveform $R^2$ of 0.986 over 2112 held-out traces, with a 5th percentile of 0.884. The apparent decrease of $R^2$ for the largest peak fields should not be interpreted as an amplitude effect, because the true ZHAireS amplitude is imposed in this test. These high-field samples are concentrated near the empirical Cherenkov-ring region, where the pulses are narrower and more sensitive to local morphology variations. The corresponding median $R^2$ after projection onto the true peak-field direction is 0.988; the projected score uses the scalar analogue of eq.~\eqref{eq:vector_waveform_r2}.

\subsection{Voltage-Domain Trigger Evaluation}

The generated electric-field trace can be propagated through a detector-response chain,
\begin{equation}
\label{eq:rfchain_mapping}
\mathbf{V}_{\mathrm{ADC}}(t,\mathbf{r})
  = \mathcal{R}_{\mathrm{RF}}
    \left[\mathbf{E}^{\mathrm{F}}(t,\mathbf{r});\theta,\phi\right].
\end{equation}
Here $\mathcal{R}_{\mathrm{RF}}$ denotes the antenna and electronics response implemented in the GRANDlib simulation pipeline~\cite{refGRANDlib}. The same receiver-chain description reproduced 24-hour Galactic-noise reception measurements within $0.7\,\mathrm{dB}$ in a self-triggered radio-detector noise and RFI study~\cite{refNoiseRFI}. This result supports the detector-response and noise-reception model but does not validate the FARSim shower waveform or array-trigger efficiency. The response operator is modular and can be replaced by an experiment-specific antenna, electronics, filtering, and digitization model without changing the FARSim field-reconstruction stage.

For the trigger-rate calculation, the ADC waveform is conditioned by a $39\,\mathrm{MHz}$ IIR notch filter followed by an FIR low-pass filter with a $115\,\mathrm{MHz}$ passband edge and a $120\,\mathrm{MHz}$ stopband edge. The local threshold-crossing logic is applied to this filtered ADC trace. In the deterministic voltage-threshold mode, the scalar trigger map is
\begin{equation}
\label{eq:voltage_peak_map}
V_{\mathrm{pk}}(\mathbf{r})
  = \max_{t,c}\left|V_{\mathrm{ADC},c}(t,\mathbf{r})\right|,
\end{equation}
and the trigger contour is extracted from $V_{\mathrm{pk}}(\mathbf{r})$ in the same way as the field-domain contour is extracted from $|\mathbf{E}_{\mathrm{pk}}(\mathbf{r})|$. This mode remains computationally efficient because the voltage map can be evaluated on a coarse grid and locally refined near the threshold-crossing region.

For a stochastic waveform-level trigger, the ADC signal is combined with a precomputed noise library and tested with level-1 logic based on two thresholds, crossing multiplicity, and quiet-time requirements, following the AERA station-trigger formulation~\cite{refAERAL1}. The trigger probability then depends on both the time series and the sampled noise realization, so a deterministic contour is no longer exact. FARSim separates the deterministic signal from reusable noise traces: the signal scales with primary energy according to eq.~\eqref{eq:energy_scaling}, whereas noise is sampled and added only at the voltage-trigger stage. This factorization avoids repeated RF-chain propagation across the energy grid and confines stochastic sampling to the final trigger decision.

Appendix~\ref{app:waveform_rfchain_diagnostics} compares the $75\,\mu\mathrm{V/m}$ field-domain contour with a deterministic voltage contour after RF-chain propagation. The array-level calculation below uses the filtered $y$-channel ADC trace.

\begin{figure}[!t]
  \centering
  \includegraphics[width=0.72\textwidth]{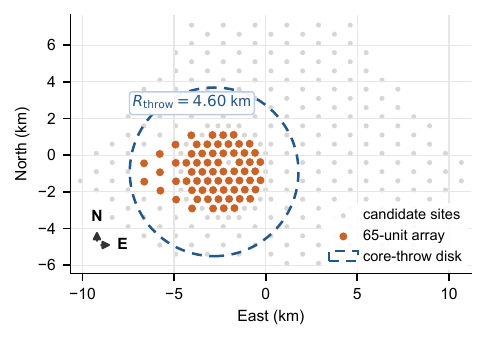}\\[0.55em]
  \includegraphics[width=0.98\textwidth]{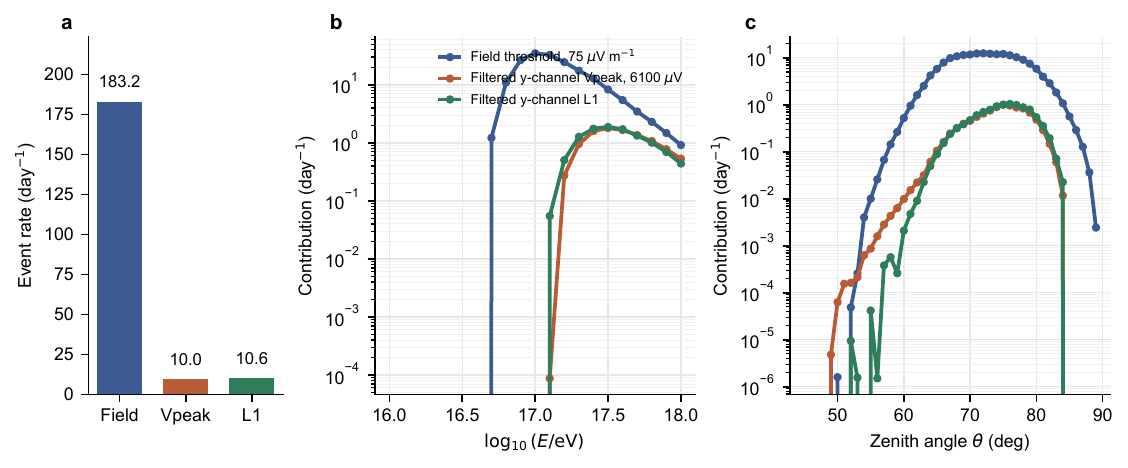}
  \caption{Diagnostic 65-unit array layout and screening-rate comparison for the three trigger abstractions considered in this RF-chain extension. The upper panel shows the 65-unit array, the surrounding candidate sites, and the fixed core-throw disk used for the rate calculation. In the lower rate panels, (a) shows the integrated screening rate for electric-field thresholding, deterministic filtered-voltage thresholding, and filtered noisy-L1 triggering, (b) shows the contribution as a function of primary energy, and (c) shows the contribution as a function of zenith angle. The voltage-domain results use the filtered $y$-channel ADC trace after one $39\,\mathrm{MHz}$ IIR notch-filter stage and an FIR low-pass stage with $f_{\mathrm{pass}}=115\,\mathrm{MHz}$ and $f_{\mathrm{stop}}=120\,\mathrm{MHz}$. The L1 parameter values are defined in the main text.}
  \label{fig:trigger_mode_rates}
\end{figure}

Figure~\ref{fig:trigger_mode_rates} compares three trigger abstractions for a 65-unit array: electric-field thresholding, deterministic filtered-voltage thresholding, and stochastic L1 triggering. The voltage threshold is $6100\,\mu\mathrm{V}$, approximately five times the RMS of the filtered noise-only $y$-channel traces. The L1 settings are $T_1=6100\,\mu\mathrm{V}$, $T_2=5500\,\mu\mathrm{V}$, $N_C=1$--10, $t_{\mathrm{prev}}=100\,\mathrm{ns}$, $t_{\mathrm{per}}=100\,\mathrm{ns}$, and $t_{C,\max}=25\,\mathrm{ns}$. Here $T_1$ and $T_2$ are the upper and lower crossing thresholds, $N_C$ is the allowed crossing multiplicity, $t_{\mathrm{prev}}$ and $t_{\mathrm{per}}$ define the quiet-time windows, and $t_{C,\max}$ is the maximum crossing separation. The integrated screening rates are 183.17, 10.01, and 10.64 events per day, respectively.

The scan uses 45 zenith-indexed reference templates. FARSim evaluates 810 direction combinations at $10^{17.5}\,\mathrm{eV}$, comprising 45 zenith angles from $45^\circ$ to $89^\circ$ and 18 azimuths in $20^\circ$ steps. Scaling these footprints to 21 energies from $10^{16}$ to $10^{18}\,\mathrm{eV}$ gives 17010 deterministic parameter-grid footprints. Each grid point is tested at 5000 core positions, corresponding to approximately $8.5\times10^7$ core placements. Bins above $85^\circ$ are retained only to display boundary behavior and do not extend the validated range.

The filtered-voltage and stochastic-L1 scan required $2.51\times10^3\,\mathrm{s}$ of wall-clock time with eight worker processes, or about 42 minutes and an order of a few CPU-hours. The separate field-threshold baseline used a denser $10^\circ$ azimuth grid and required approximately $96\,\mathrm{s}$ with four workers. These rates compare trigger abstractions under a fixed synthetic-noise prescription; accidental triggers, dead time, station-to-station calibration, and site-specific noise have not been validated for a deployed detector.

\begin{figure}[H]
  \centering
  \includegraphics[width=0.82\textwidth]{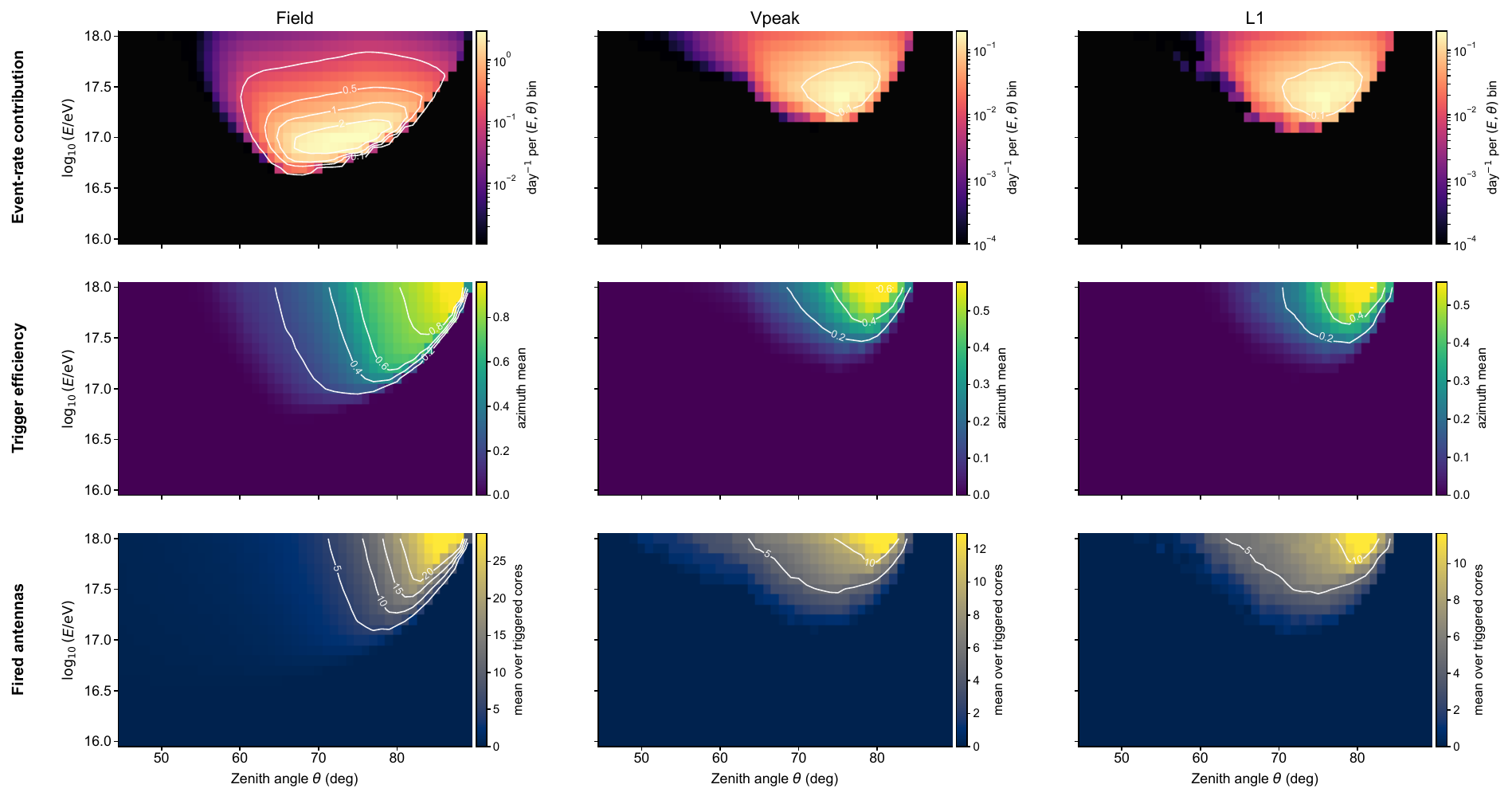}
  \caption{Binned 65-unit array diagnostics for the same three trigger abstractions as in figure~\ref{fig:trigger_mode_rates}. The columns correspond to electric-field thresholding, deterministic filtered-voltage thresholding, and filtered noisy-L1 triggering. The rows show the event-rate contribution, trigger efficiency, and the mean number of triggered antennas among triggered events. The voltage-domain panels use the filtered $y$-channel ADC trace and the same $6100\,\mu\mathrm{V}$ threshold scale as the rate profiles.}
  \label{fig:trigger_mode_maps_filtered}
\end{figure}

The corresponding binned event-rate, trigger-efficiency, and station-multiplicity maps are shown in figure~\ref{fig:trigger_mode_maps_filtered}.

\clearpage

\FloatBarrier

\subsection{Scope of the Extension}

The waveform extension connects fast footprint reconstruction to detector-response studies but does not replace microscopic simulation for precision waveform reconstruction, composition inference, or electronics validation. It supports field-threshold contours, deterministic voltage thresholds, and stochastic waveform-level L1 tests within the same footprint model. The latter adds detector-response and noise sensitivity while remaining a screening calculation rather than a validated detector trigger.

\section{Conclusions and Outlook}
\label{sec:conclusion}

FARSim is a trigger-screening surrogate that combines geomagnetic and charge-excess field reconstruction, geomagnetic-angle and energy scaling, geometrical projection, and contour-based core sampling. Across 941 retained events from an independent 1999-shower validation set, the reconstructed geomagnetic lateral profiles have a median relative RMSE of 9.4\% and a median Pearson coefficient of $r_{\mathrm P}=0.965$. The pulse-shape extension reaches a median vector-waveform $R^2$ of 0.986 on 2112 traces held out from pulse-shape model fitting when the true peak amplitude is imposed.

The method is validated up to $85^\circ$, with $60^\circ$--$85^\circ$ as the principal range for sparse arrays with station spacings of order $500\,\mathrm{m}$. More vertical showers require denser sampling and, where relevant, clipping-aware references. Above $85^\circ$, radio-core recentering does not capture all event-dependent source-distance, coherence, polarization, and two-dimensional footprint effects.

Field contours provide the lowest-cost screening mode, while deterministic voltage thresholds and stochastic L1 tests expose the sensitivity of trigger regions to detector response and noise. The reported rates compare these abstractions but are not exposure predictions for a deployed instrument. Extending the model will require broader $X_{\max}$ and waveform coverage, site-specific noise and calibration, accidental-rate and dead-time estimates, and validation against full detector simulations and data.

\clearpage
\appendix

\section{Auxiliary Method Figures}
\begin{figure}[H]
  \centering
  \subfloat[]{\includegraphics[width=0.42\textwidth]{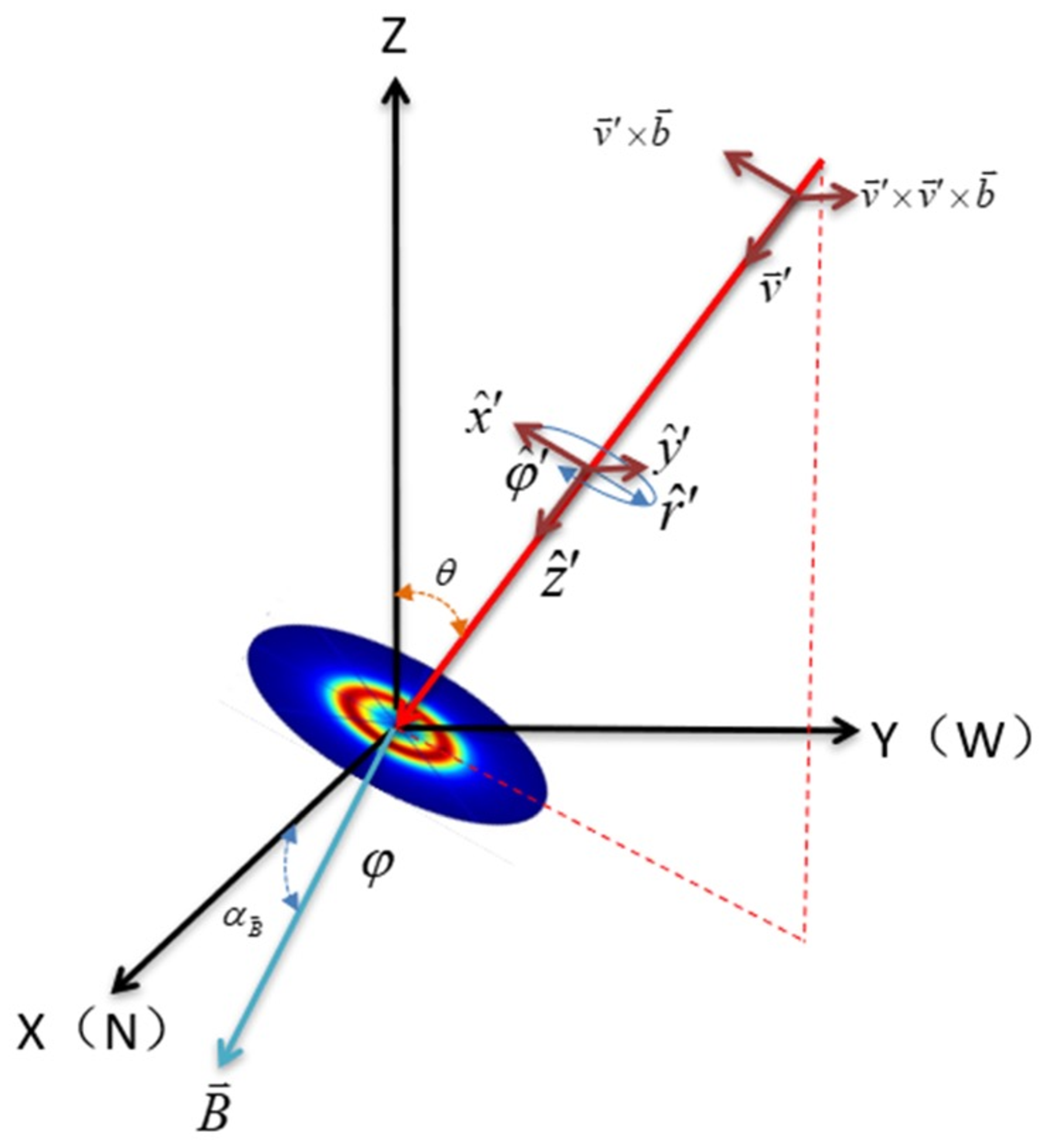}}
  \hfill
  \subfloat[]{\includegraphics[width=0.42\textwidth]{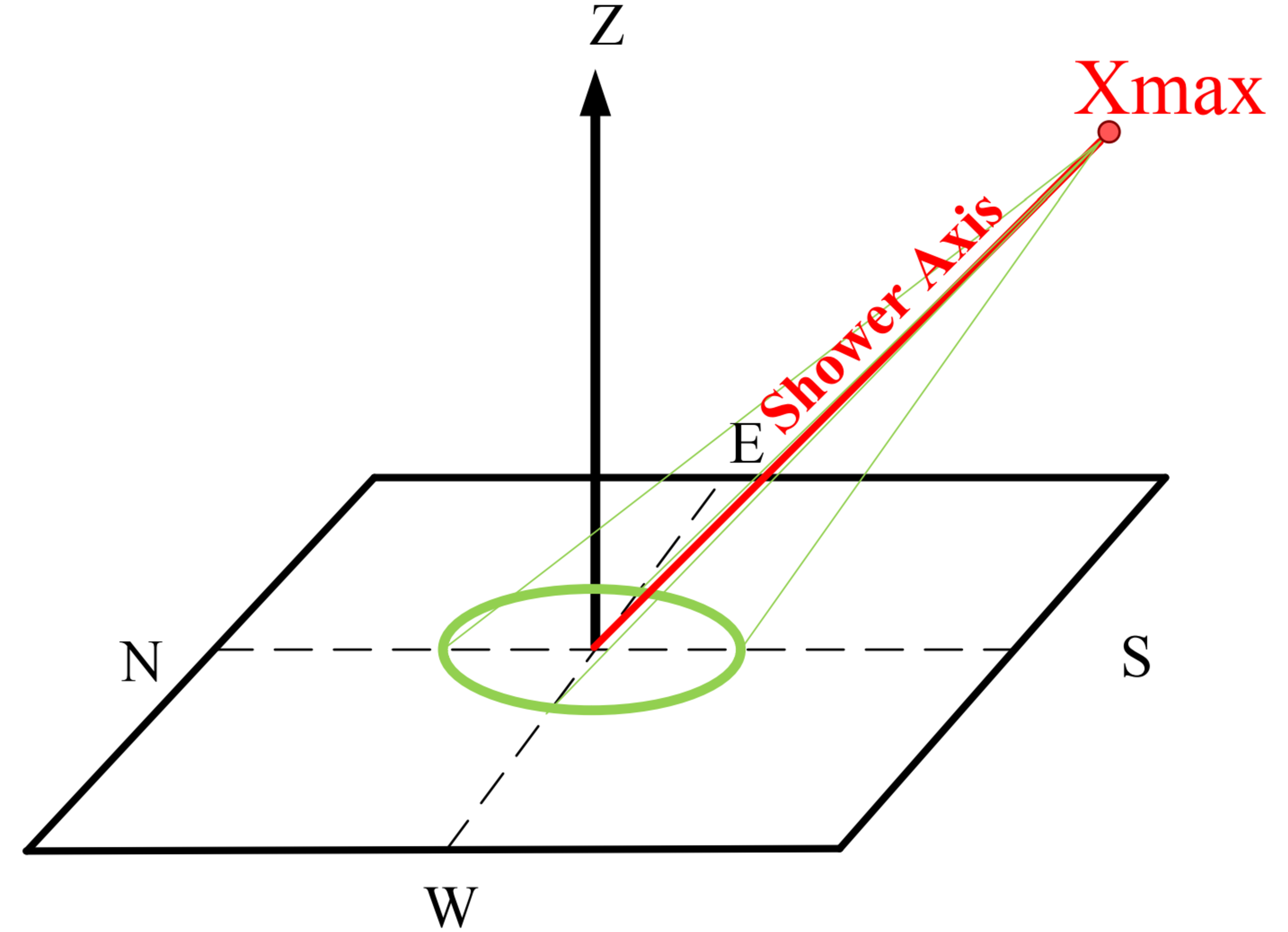}}\\[-0.3ex]
  \caption{Diagram of coordinate systems: (a) transformation from the global Cartesian system to shower-based coordinates $(x',y',z')$, (b) geometry of the EAS shower axis and $X_{\max}$.}
  \label{fig:coordinate_diagram}
\end{figure}

\begin{figure}[H]
  \centering
  \includegraphics[width=0.90\textwidth,height=0.29\textheight,keepaspectratio]{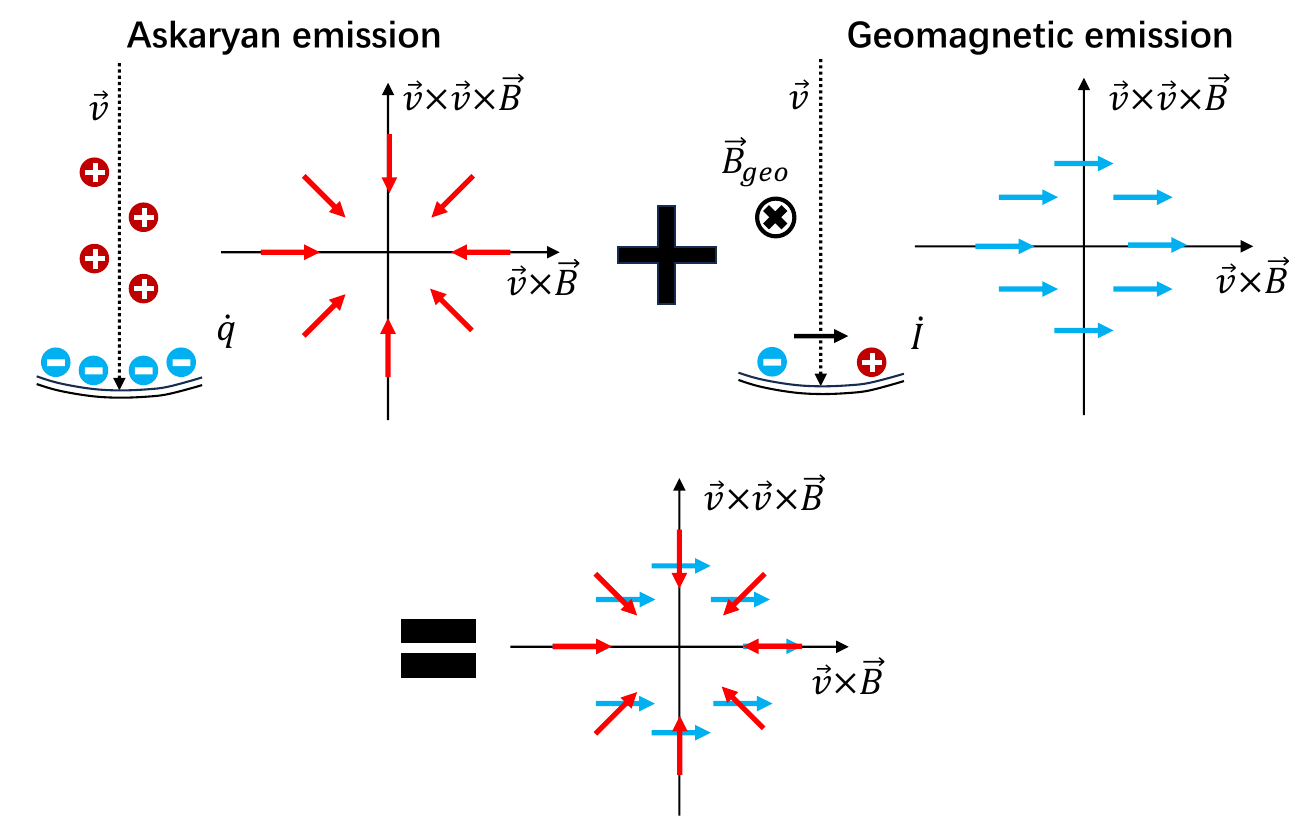}
  \caption{Schematic of the geomagnetic and charge-excess emission mechanisms contributing to the total field $\vec{E}_T$.}
  \label{fig:geomagnetic_askaryan}
\end{figure}

\begin{figure}[H]
  \centering
  \includegraphics[width=0.95\textwidth,height=0.28\textheight,keepaspectratio]{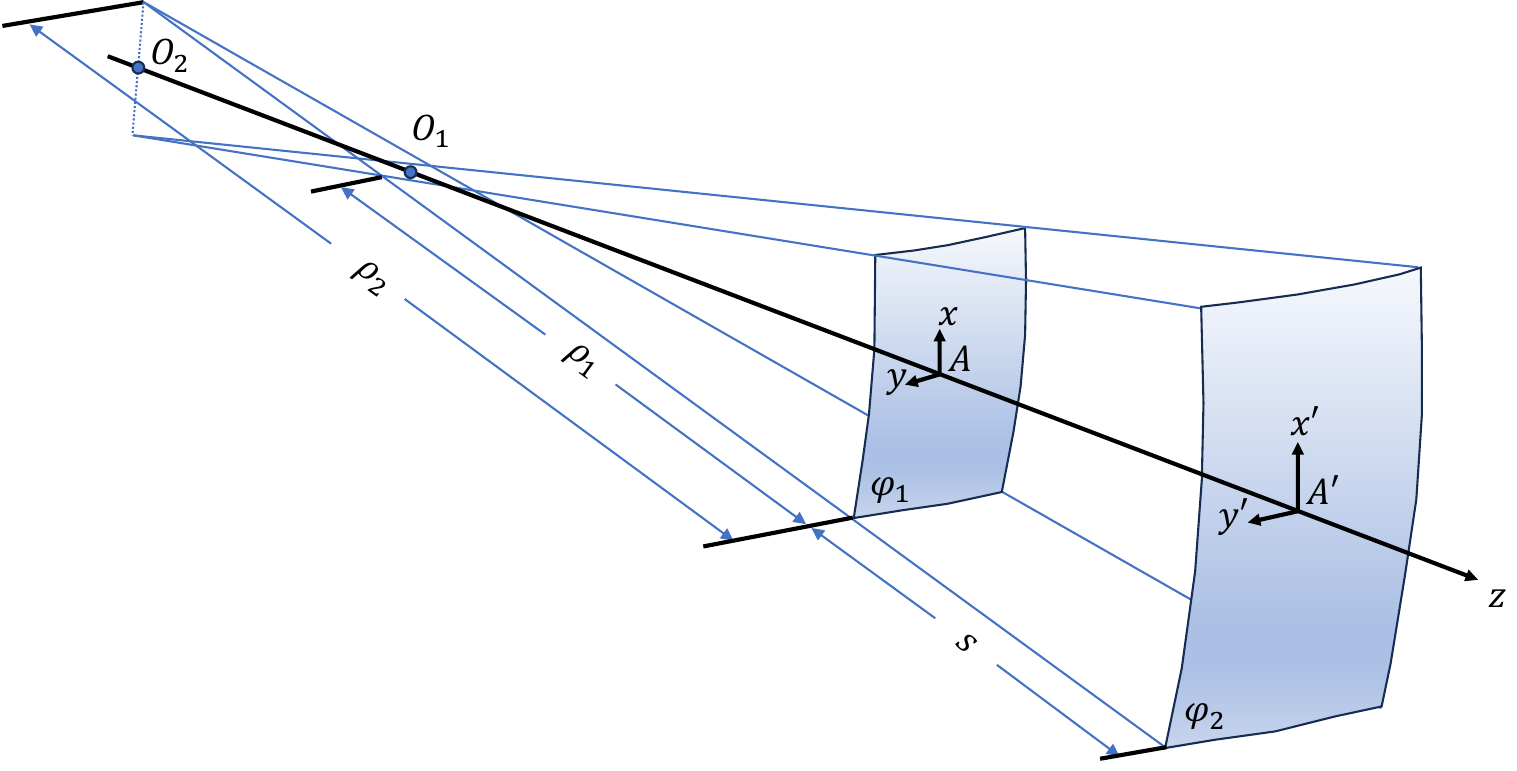}
  \caption{Geometrical projection of the electric field from one wavefront surface to another. The notation corresponds to eq.~\eqref{eq:field_projection}, where the projection distance $s$ and the wavefront-curvature radii $\rho_1$ and $\rho_2$ determine the field-amplitude correction.}
  \label{fig:projection_geometry}
\end{figure}

\begin{figure}[H]
  \centering
  \includegraphics[width=0.78\textwidth,height=0.44\textheight,keepaspectratio]{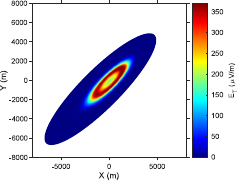}
  \caption{Projected ground-plane field distribution obtained after geometrical transformation from the shower plane.}
  \label{fig:projection_method}
\end{figure}

Figures~\ref{fig:coordinate_diagram}, \ref{fig:geomagnetic_askaryan}, \ref{fig:projection_geometry}, and~\ref{fig:projection_method} collect method-definition diagrams and illustrative footprint transformations that support the FARSim construction but are not required for the main validation argument.

\clearpage

\clearpage

\section{Waveform and RF-Chain Diagnostic Figures}
\label{app:waveform_rfchain_diagnostics}

\begin{figure}[H]
  \centering
  \includegraphics[trim=19bp 6bp 0bp 0bp,clip,width=\textwidth]{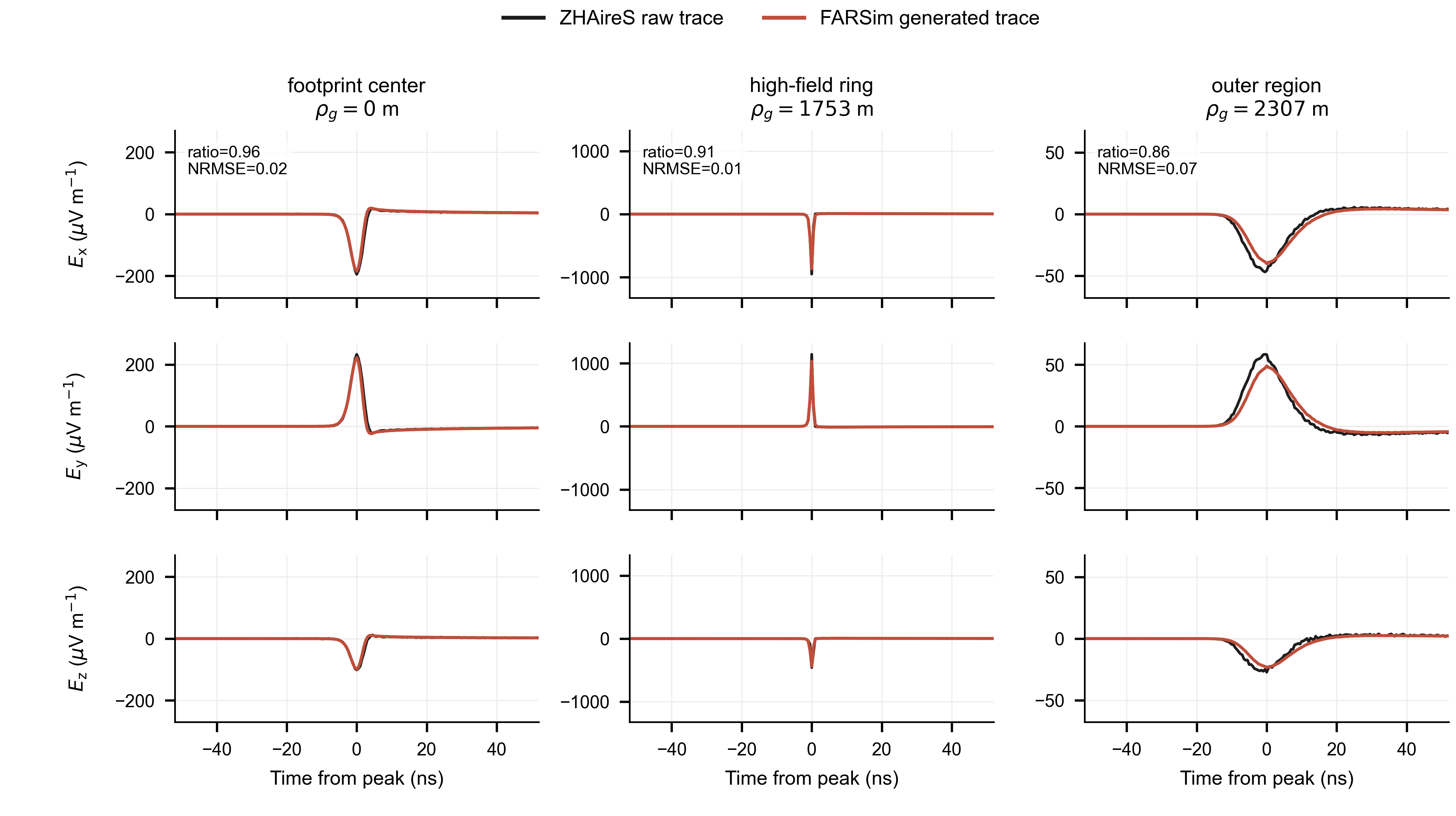}
  \caption{Time-domain electric-field traces generated by FARSim and compared with raw ZHAireS traces for three representative station positions in one inclined iron-shower event. The columns correspond to the footprint center, a high-field ring station, and an outer-region station; the rows show the three Cartesian field components. The traces are aligned at the vector-field peak of the ZHAireS trace. The annotated ratio is the FARSim-to-ZHAireS vector peak ratio, and NRMSE is evaluated over the displayed time window.}
  \label{fig:farsim_waveform_examples}
\end{figure}

\begin{figure}[H]
  \centering
  \includegraphics[width=\textwidth]{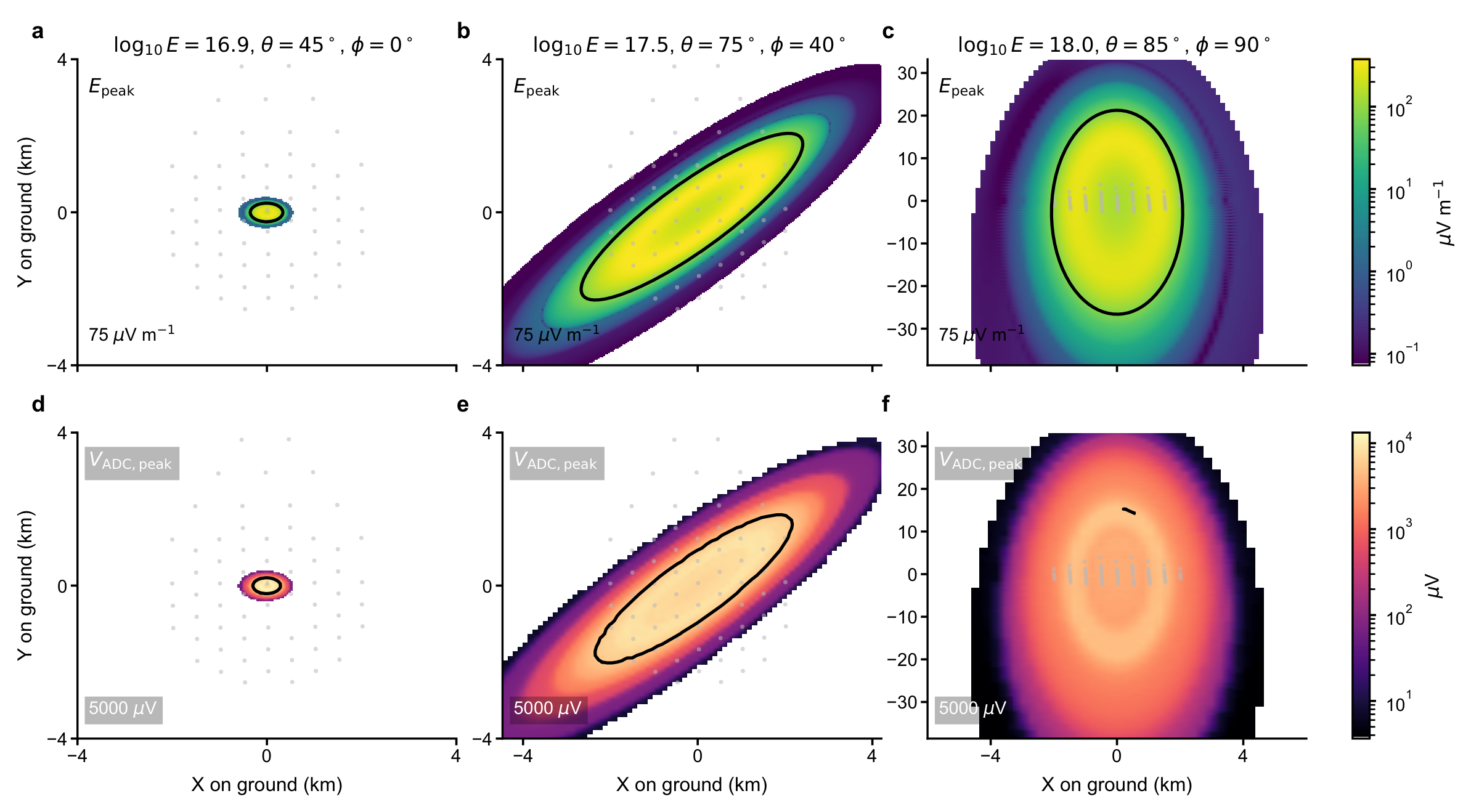}
  \caption{Comparison of field-domain and voltage-domain trigger regions for three representative showers. The top row shows FARSim ground-plane electric-field maps with the $75\,\mu\mathrm{V/m}$ contour, and the bottom row shows the signal-only RF-chain output maps with the $5000\,\mu\mathrm{V}$ ADC threshold contour. The voltage maps are evaluated only inside the electric-field support region to avoid artificial extrapolated voltage footprints. The extracted field-contour areas are 0.264, 8.522, and $156.293\,\mathrm{km^2}$, whereas the corresponding voltage-contour areas are 0.188, 6.723, and $0.003\,\mathrm{km^2}$.}
  \label{fig:rfchain_trigger_maps}
\end{figure}

Figures~\ref{fig:farsim_waveform_examples} and~\ref{fig:rfchain_trigger_maps} provide the station-level waveform and footprint-level RF-chain diagnostics supporting section~\ref{sec:waveform_rfchain}.

\clearpage

\FloatBarrier

\acknowledgments
The authors would like to thank Olivier Martineau-Huynh for his valuable comments and suggestions, which helped improve the validation strategy and the presentation of this work. This work was supported by the National Square Kilometre Array (SKA) Program of China under Grant 2025SKA0110101.

\section*{Data, Software and Code Availability}

The ZHAireS input files, processed reference-footprint tables, trained waveform-template parameters, plotting data, and FARSim analysis scripts used to produce the results in this article are available from the corresponding author upon reasonable request. The raw full-trace ZHAireS libraries are not deposited with the article because of their large file size, but the processed footprints and summary tables needed to reproduce the validation figures and screening-rate calculations can be shared. External simulation and detector-response tools used in the workflow are cited in the text.

\end{document}